\documentclass[10pt,final,journal,a4paper,oneside,twocolumn]{IEEEtran}
\usepackage{cite}

\usepackage{amsmath,amsfonts,amssymb}
\usepackage{mathdots}

\makeatletter
\let\MYcaption\@makecaption
\makeatother

\usepackage[font=footnotesize]{subcaption}

\makeatletter
\let\@makecaption\MYcaption
\makeatother

\usepackage{bbm}
\usepackage{mathtools, cuted}
\usepackage{amssymb,bm}
\usepackage{hyperref}
\usepackage{ragged2e}
\usepackage{overpic}
\usepackage{float}
\usepackage{color}
\usepackage[super]{nth}
\usepackage{graphicx}
\usepackage{textcomp}
\usepackage{tikz}
\usetikzlibrary{calc,positioning}
\usepackage{array}
\usepackage[crop=pdfcrop]{pstool}
\usepackage{cancel}
\usepackage{dblfloatfix} % Table at bottom of page
\usepackage{tabularx}
\usepackage[usestackEOL]{stackengine}
\usepackage{blindtext}
\usepackage{float}
\usepackage{siunitx}

\newcommand{\htext}[1]{%
	\makebox[0pt]{\Centerstack{#1}}
}

\newcommand{\vtext}[1]{%
	\makebox[0pt]{\rotatebox[origin=c]{90}{\Centerstack{#1}}}
}

\newcommand{\st}{\text{s}}
\newcommand{\ptt}{\text{p}}
\newcommand{\St}{\text{S}}
\newcommand{\Tt}{\text{T}}

\newcommand{\Ft}{\text{F}}
\newcommand{\TQt}{\text{TQ}}

\newcommand{\NCt}{\text{NC}}
\newcommand{\ntt}{\text{n}}

\newcommand{\bbmatrix}{\begin{bmatrix}}
\newcommand{\ebmatrix}{\end{bmatrix}}

\newcommand{\E}{\mathbf{E}}

\newcommand{\C}{\mathbf{C}}
\renewcommand{\H}{\mathbf{H}}

\newcommand{\NN}{\mathbb{\bar N}}
\newcommand{\J}{\mathbf{J}}
\newcommand{\K}{\mathbf{K}}

\newcommand{\M}{\mathbf{M}}
\renewcommand{\P}{\mathbf{P}}
\renewcommand{\r}{\mathbf{r}}
\newcommand{\V}{\mathbf{V}}
\newcommand{\F}{\mathbf{F}}

\renewcommand{\P}{\mathbf{P}}

\newcommand{\rb}{\boldsymbol{\r}}
\newcommand{\rbb}{\mathbbm{r}}
\renewcommand{\L}{\mathbf{\mathcal{L}}}
\newcommand{\R}{\mathbf{\mathcal{R} }}
\newcommand{\Nh}{\mathbf{\hat n}}
\newcommand{\Nu}{\mathbf{N}}

\newcommand{\0}{\varnothing}
\newcommand{\I}{\mathbb{I}}

\newcommand{\Cv}{\mathbb{C}}

\newcommand{\Pv}{\overline{\mathbb{P}}}

\newcommand{\Pvbb}{\mathbb{P}}
\newcommand{\Sv}{\mathbb{S}}

\newcommand{\Rv}{\mathbb{R}}

\newcommand{\Lv}{\mathbb{L}}
\newcommand{\Ev}{\mathbb{E}}

\newcommand{\Dv}{\mathbb{D}}
\newcommand{\Gv}{\mathbb{G}}
\newcommand{\Jv}{\mathbb{J}}
\newcommand{\Mv}{\mathbb{M}}
\newcommand{\Hv}{\mathbb{H}}

\newcommand{\Kv}{\mathbb{K}}

\newcommand{\chit}{\overline{\overline{\chi}}}

\newcommand{\Q}{\mathbf{Q}}
\newcommand{\El}{\boldsymbol{\mathcal{E}}}
\newcommand{\Nl}{\boldsymbol{\mathcal{N}}}

\newcommand{\Ql}{\boldsymbol{\mathcal{Q}}}
\newcommand{\Rlx}{\boldsymbol{\mathcal{R_\times}}}
\newcommand{\Hl}{\boldsymbol{\mathcal{H}}}
\newcommand{\Pln}{\mathcal{P}_n}
\newcommand{\Ml}{\boldsymbol{\mathcal{M}}}
\newcommand{\Pl}{\boldsymbol{\mathcal{P}}}
\newcommand{\Mln}{\mathcal{M}_n}

\newcommand{\nh}{\mathbf{\hat n}}
\newcommand{\nhl}{\boldsymbol{\mathcal{\hat N}}}
\newcommand{\lh}{\boldsymbol{\hat \ell}}
\newcommand{\rh}{\mathbf{\hat r}}
\renewcommand{\th}{\mathbf{\hat t}}
 \newcommand{\Vl}{\boldsymbol{\mathcal{V}}}
\newcommand{\chil}{\boldsymbol{{\mathcal{X}}}}
\newcommand{\XlT}{\boldsymbol{X^\text{T}}}

\newcommand{\Xln}{\boldsymbol{{X}^\text{n}}}
\newcommand{\Nv}{\mathbb{N}}
\newcommand{\xh}{\mathbf{\hat x}}
\newcommand{\yh}{\mathbf{\hat y}}
\newcommand{\zh}{\mathbf{\hat z}}

\newcommand{\Vf}{\boldsymbol{\mathfrak{V}}}
\newcommand{\Ef}{\boldsymbol{\mathfrak{E}}}
\newcommand{\Hf}{\boldsymbol{\mathfrak{H}}}
\newcommand{\Xf}{\boldsymbol{\mathfrak{X}}}
\newcommand{\Sf}{\boldsymbol{\mathfrak{S}}}
\newcommand{\Nf}{\boldsymbol{\mathfrak{N}}}
\newcommand{\Ff}{\boldsymbol{\mathfrak{F}}}
\newcommand{\Df}{\boldsymbol{\mathfrak{D}}}
\newcommand{\Gf}{\boldsymbol{\mathfrak{G}}}

\newcommand{\Lf}{\boldsymbol{\mathfrak{L}}}

\newcommand{\ee}{\text{ee}}
\newcommand{\zz}{{zz}}
\newcommand{\zt}{{zt}}
\newcommand{\zn}{{zn}}
\newcommand{\nn}{{nn}}  
  
\newcommand{\nt}{{nt}}
\newcommand{\nz}{{nz}}
\newcommand{\tn}{{tn}}
\renewcommand{\tt}{{tt}}
\newcommand{\tz}{{tz}}
\newcommand{\av}{\text{av}}

\newcommand{\mm}{\text{mm}}
\newcommand{\me}{\text{me}}
\newcommand{\emm}{\text{em}}

\definecolor{burntorange}{rgb}{0.8, 0.28, 0.0}
\definecolor{myGreen}{rgb}{0.0, 0.5, 0.0}
\definecolor{amber}{rgb}{0.8, 0.28, 0.0}
\definecolor{ceruleanblue}{rgb}{0.16, 0.28, 0.75}

\usepackage[normalem]{ulem}

\newcommand{\chia}[2]{\chi_{\text{#1}}^{#2}}

\definecolor{ao}{rgb}{0.0, 0.5, 0.0}
\definecolor{cobalt}{rgb}{0.0, 0.28, 0.67}
\definecolor{amber}{rgb}{0.8, 0.36, 0.27}
\usepackage{balance}
\begin{document}

\title{IE-GSTC Metasurface Field Solver using Surface Susceptibility Tensors with Normal Polarizabilities}

\author{Tom. J. Smy, Ville Tiukuvaara \IEEEmembership{Student Member, IEEE}, \\ and Shulabh Gupta \IEEEmembership{Senior Member, IEEE}
\thanks{ Tom J. Smy, Ville Tiukuvaara, and Shulabh Gupta are with Carleton University, Ottawa, Canada (e-mail: tjs@doe.carleton.ca). }}

\maketitle

\begin{abstract}
An Integral Equation (IE) based electromagnetic field solver using metasurface susceptibility tensors is proposed and validated using variety of numerical examples in 2D. The proposed method solves for fields generated by the metasurface which are represented as spatial discontinuities satisfying the Generalized Sheet Transition Conditions (GSTCs), and described using tensorial surface susceptibility densities, $\bar{\bar{\chi}}$. For the first time, the complete tensorial representation of susceptibilities is incorporated in this integrated IE-GSTC framework, where the normal surface polarizabilities and their spatial derivatives along the metasurface are rigorously taken into account. The proposed field equation formulation further utilizes a local co-ordinate system which enables modeling metasurfaces with arbitrary orientations and geometries. The proposed 2D BEM-GSTC framework is successfully tested using variety of examples including infinite and finite sized metasurfaces, periodic metasurfaces and complex shaped structures, showing comparisons with both analytical results and a commercial full-wave solver. It is shown that the zero-thickness sheet model with complete tensorial susceptibilities can very accurately reproduce the macroscopic fields, accounting for their angular field scattering response and the edge diffraction effects in finite-sized surfaces.
\end{abstract}


%\tableofcontents

\section{Introduction}

Electromagnetic (EM) Metasurfaces, which are 2D counterparts of electromagnetic metamaterials, have recently become a topic of intense research efforts due to their versatile wave transformation capabilities and the ease of creating practical implementations across the EM spectrum. They are based on 2D arrangements of sub-wavelength resonating particles of various geometrical shapes, sizes and material constructs \cite{MS_review_Yu, Chi_Review}. Engineering these resonating particles at a microscopic scale and their specific arrangements along a surface, determines their macroscopic interactions with the EM fields around them \cite{GSTC_Holloway, meta2}.

Besides the requirements of innovating novel resonating particles and corresponding effects, an equally important problem in metasurface engineering is an efficient EM modeling and numerical computation of the fields scattered off them in response to  specified excitation fields. Field scattering from metasurfaces is inherently a \emph{multi-scale problem}, ranging from microscopic resonators to their electrically large 2D arrangement forming an array. Consequently, brute force simulations of 2D arrays of 3D geometrical resonators using standard commercial software is not the most practical choice. 

To reduce the computational complexity and speed, metasurface resonators can be described using spatially varying electric and magnetic surface susceptibility models accounting for dipolar interactions, $\bar{\bar{\chi}}_\text{e,m}(\r, \omega)$, which reduce a practical metasurface into an effective zero thickness sheet model \cite{Chi_Review, MS_Synthesis, Chi_extraction_Macrodmodel, TBC_vs_GSTC_Caloz}. The electromagnetic fields interacting with the metasurfaces, now being  represented as a spatial discontinuity, must now satisfy the Generalized Sheet Transition Conditions (GSTCs) \cite{KuesterGSTC, GSTC_Holloway}, which are the sheet counterparts of volumetric Maxwell's equations. These surface susceptibility models are typically extracted using simulations with a single volumetric resonator placed within periodic boundaries to emulate an infinite surface, and excited with a chosen set of incident fields \cite{GenBCEM}.%These surface susceptibility models are typically extracted from practical volumetric resonators in periodic unit cell environments when excited with a chosen set of incident waves \cite{GenBCEM}.

The general form of surface susceptibility tensors include electric ($\bar{\bar{\chi}}_\text{ee}$) and magnetic susceptibilities ($\bar{\bar{\chi}}_\text{mm}$) in addition to the bi-anisotropic tensors $\bar{\bar{\chi}}_\text{em}$ and $\bar{\bar{\chi}}_\text{me}$. Each of these tensors consists of 9 complex components, so that the resulting 36 susceptibility components account for various EM effects such as polarization rotation, dissipation losses, reciprocity and angular scattering properties of the surface \cite{MS_Synthesis}. Once the surface susceptibilities are known, they can next be used in conjunction with the GSTCs to efficiently solve for the macroscopic EM field scattering from a general non-uniform metasurface for arbitrary field excitations, compared to brute force 3D resonator array simulations which are much more computationally-intensive as they capture the microscopic fields within each resonator.

Metasurface field analysis typically involves integrating GSTCs into bulk Maxwell's equations using a variety of standard numerical techniques based on Finite-Difference and Finite Element methods \cite{Caloz_MS_Siijm, Caloz_Spectral, Smy_Metasurface_Space_Time}, and Integral-Equation (IE) based techniques \cite{Smy_Close_ILL, smy2020IllOpen, stewart2019scattering, FE_BEM_Impedance, Caloz_MS_IE, AppBEMEM, Smy_EuCap_BEM_2020, Caloz_EM_inversion}. For electrically large metasurface problems, IE-GSTC methods are a computationally efficient choices as only the surfaces (as opposed to volumes) are meshed and solved for equivalent electric and magnetic currents, $\J(\r)$ and $\K(\r)$, which are then used to calculate the  propagating fields at desired observation locations only.

Due to the 2D nature of metasurfaces, the surface susceptibilities are naturally decomposed into tangential and normal components. Until recently, the normal susceptibility components have been typically ignored citing problem simplicity, and metasurfaces have thus largely been analyzed using tangential components only \cite{smy2020IllOpen, MS_Synthesis, CalozFDTD}. However, with rapidly growing works on the surface susceptibility models, it has become clear that these normal components play a crucial role in accurate modeling of metasurfaces, in particular, with their angular scattering response, while the tangential components are generally suitable for paraxial wave propagation only \cite{Karim_Angular_MS, Karim_Bianiso_MS}.

All previous works on computational modeling and field scattering from metasurfaces, based on a variety of methods, have ignored the normal surface susceptibilities. While it may appear that their extension to include normal components is straightforward, in practice, its not the case as the normal components inside GSTCs involve spatial derivatives of surface polarizations and proper care must be taken in solving GSTC based field equations. In this work, a 2D IE-GSTC based metasurface field solver framework is presented, rigorously accounting for these normal surface susceptibility components in addition to the tangential ones. Using a surface susceptibility description in a local co-ordinate system, the proposed framework is applicable to arbitrary curvilinear metasurfaces with spatially varying surface susceptibilities. Moreover, the framework includes all 36 components of the susceptibility tensors, and is thereby capable of modeling and solving scattered fields from a given general metasurface.

The paper is structured as follows. Sec.~II presents the general problem of field scattering from metasurfaces and motivates the necessity to include the normal surface susceptibility components in practical metasurface problems. It further initiates the basics of the IE based field propagation framework followed by surface description in a local co-ordinate system followed later in this work. Sec.~III presents the analytical formulation of the IE-GSTC field scattering problem and describes how the spatial gradients of the normal surface susceptibilities can be computed in a local co-ordinate system. Sec.~IV translates the analytical field equations using numerical discretization based on Boundary Element Method (BEM) and presents the computational framework that is implemented next to solve for the scattered fields. Various examples of uniform and non-uniform metasurfaces are next presented in Sec.~V to illustrate the method and the importance of including normal susceptibility components along with several convergence studies. Finally, conclusions are made in Sec.~VI. A brief appendix is provided at the end presenting analytical results of a periodic metasurface based on Floquet field expansions \cite{VilleFloq} to be used for validation of some of the examples given in Sec.~V.

\section{Motivation \& Problem Description}\label{Sec:PDesc}

\subsection{Importance of Normal Surface Susceptibility Components}

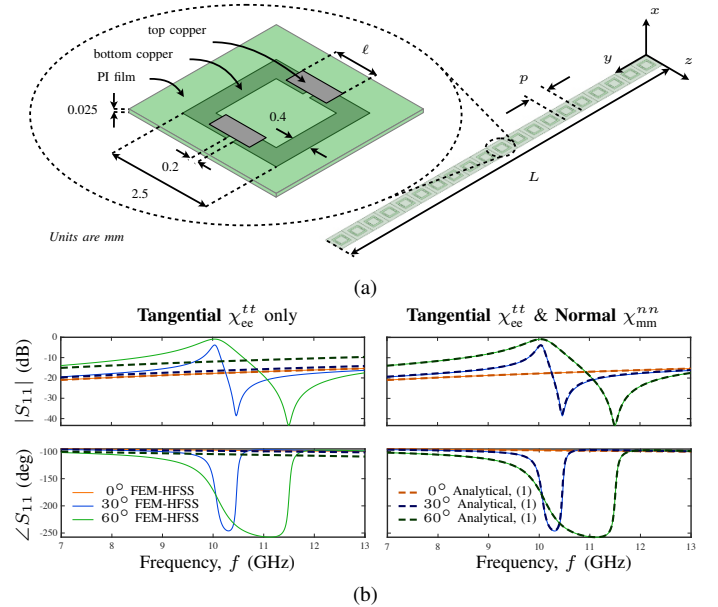
\begin{figure}[htbp]
    \centering
    \begin{subfigure}[b]{\linewidth}
       \begin{overpic}[width=1\linewidth,grid=false,trim={0cm 0cm 0cm 0cm},clip]{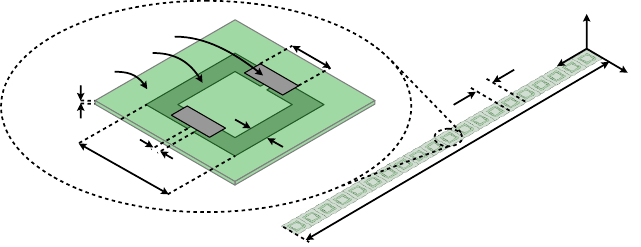}
				\put(8,22){\htext{\tiny 0.025}}
				\put(16.5,9.5){\htext{\tiny 2.5}}
				\put(15.5,30.5){\htext{\tiny bottom copper}}
				\put(22,34){\htext{\tiny top copper}}
				\put(13,27){\htext{\tiny PI film}}
				\put(21,13.5){\htext{\tiny 0.2}}
				\put(37,21){\htext{\tiny 0.4}}
				\put(50,31){\htext{\tiny $\ell$}}
				\put(3,3){\tiny \textit{Units are mm}}
				\put(73.5,27){\htext{\tiny $p$}}
				\put(75,12){\htext{\tiny $L$}}%
				\put(86,30){\htext{\tiny $y$}}
				\put(98,29.5){\htext{\tiny $z$}}
				\put(93,37){\htext{\tiny $x$}}
			\end{overpic}\caption{}
	 \end{subfigure}
 	\begin{subfigure}[b]{\linewidth}
  			  \begin{overpic}[width=1\linewidth,grid=false,trim={0cm 0cm 0cm 0cm},clip]   {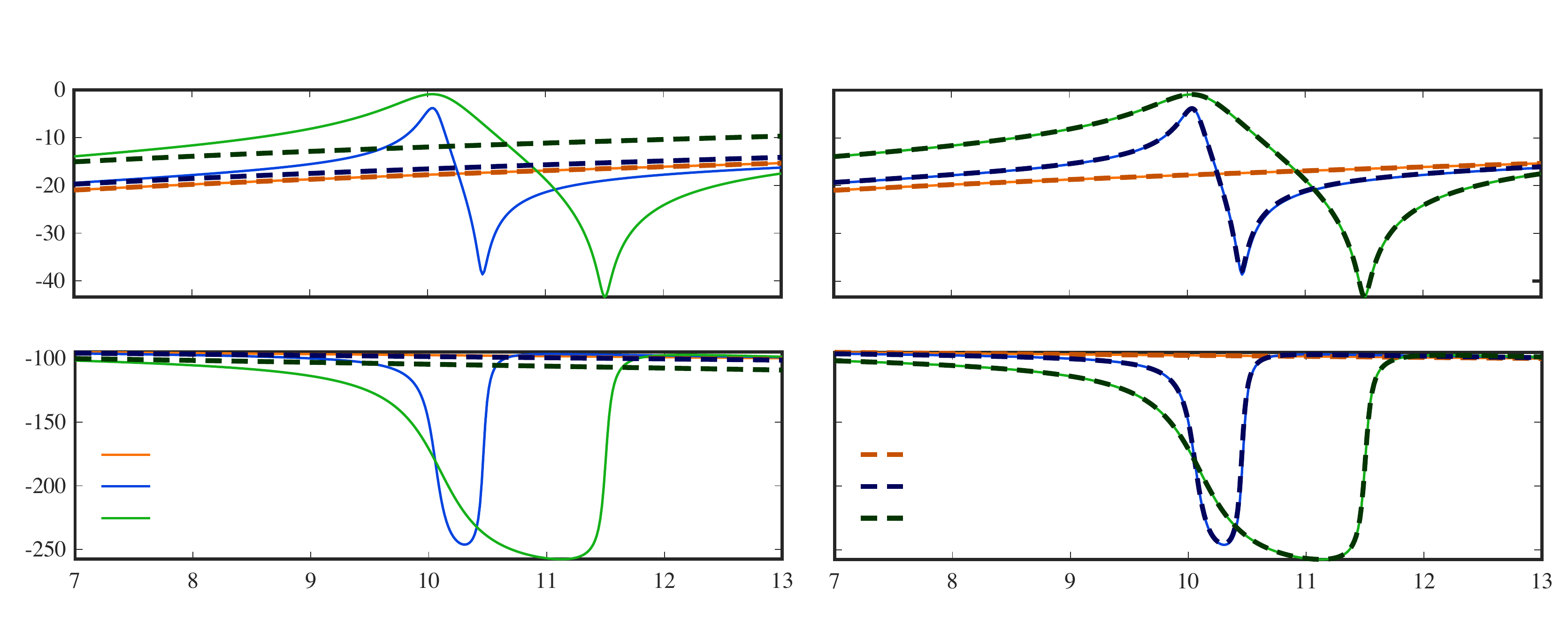}
			   \put(28, 0){\htext{\scriptsize Frequency, $f$~(GHz)}}
			    \put(78, 0){\htext{\scriptsize Frequency, $f$~(GHz)}}
			     \put(-1, 10){\vtext{\scriptsize $\angle S_{11}$~(deg)}}
			       \put(-1, 27){\vtext{\scriptsize $|S_{11}|$~(dB)}}
			        \put(28, 37){\htext{\scriptsize \textbf{Tangential} $\chi_\text{ee}^\tt$ only}}
			 \put(75, 37){\htext{\scriptsize \textbf{Tangential} $\chi_\text{ee}^\tt$ \& \textbf{Normal} $\chi_\text{mm}^\nn$}}
			 \put(18, 11){\htext{\tiny $0^\circ$ FEM-HFSS}}
			  \put(18, 9 ){\htext{\tiny $30^\circ$ FEM-HFSS}}
			   \put(18,  7){\htext{\tiny $60^\circ$ FEM-HFSS}}
			    \put(67, 11){\htext{\tiny $0^\circ$ Analytical, \eqref{eq:AnaTR}}}
			     \put(67, 9){\htext{\tiny $30^\circ$ Analytical, \eqref{eq:AnaTR}}}
			      \put(67, 7){\htext{\tiny $60^\circ$ Analytical, \eqref{eq:AnaTR}}}
        \end{overpic}\caption{}
    \end{subfigure}
        	\caption{An example of a practical metasurface composed of sub-wavelength capacitively-loaded loops on a Polyimide (PI) film ($\epsilon_r=3.4$, loss tangent 0.002) exhibiting a strong magnetic dipolar response normal to the surfaces. a) The metasurface schematic and unit cell description. b) FEM-HFSS computed reflection response for varying angles of plane-wave incidences. The surface is composed of 20 cells having a side length $p=\SI{4}{mm}$. TE polarization and angular scan in the $x$-$z$ plane is assumed.}\label{fig:PracticalCell}
\end{figure}

Consider an illustrative example of a practical uniform metasurface (in the $y$-$z$ plane) consisting of split metallic loops on a thin Polyimide film as shown, as shown in Fig.~\ref{fig:PracticalCell}(a). The loops are loaded with metal-insulator-metal (MIM) capacitors, which can control their resonant frequencies, and the metasurface is excited with $z-$polarized incident fields. The typical reflection characteristics of this structure are shown in Fig.~\ref{fig:PracticalCell}(b) computed using Ansys FEM-HFSS (transmission not shown for brevity). While a weak reflection is observed under normal plane-wave incidence, a very strong reflection indicative of significant resonator interaction with the wave is observed around 10~GHz under oblique incidence conditions (at 30$^\circ$ and $60^\circ$, for example). Let us see how a surface susceptibility based zero thickness sheet may be used to model this metasurface response.

Under normal plane-wave incidence, parallel electric currents can be induced in the MIM capacitor branches forming an electric resonance. This can be modeled using a \emph{tangential} ($z-$polarized) electric surface susceptibilities, $\chi_\text{ee}^\tt$. When instead an oblique plane-wave is incident in the $x-y$ plane, a time-varying magnetic flux through each loop generates a circulating current resulting in a strong $x-$polarized magnetic dipolar moment. This is modeled using a \emph{normal} magnetic surface susceptibility, $\chi_\text{mm}^\nn$ ($x-$polarized). By symmetry and reciprocity considerations, it can be shown that these are the only two dominant susceptibility components describing the electromagnetic interaction of this unit cell with incident waves \cite{Karim_Angular_MS}. Simulations were undertaken of the unit cells with periodic boundary conditions and using plane-wave incidence at two different angles as prescribed in \cite{Xiao:2019aa,VilleFloq} to extract these susceptibilities as a function of frequency, $\omega$. 

An analysis of plane wave propagation onto a flat uniform metasurface of infinite extent described by tangential $\chi_\text{ee}^\tt$ and normal $\chi_\text{mm}^\nn$, at an arbitrary angle of incidence of $\theta$ can provide an angle-dependent reflection and transmission \cite{GenBCEM}, given by

\begin{subequations} \label{eq:AnaTR}
    \begin{align}
    R(\theta, \omega) &= \left[\frac{-j k_0(\chi_\ee^\tt  + \chi_\mm^\nn \sin^2\theta)}{2\cos\theta -  jk_0  (\chi_\ee^\tt + \chi_\mm^\nn\sin^2\theta)} \right]\\
    T(\theta, \omega) &= \left[\frac{2\cos\theta}{2\cos\theta + jk_0 (\chi_\ee^\tt + \chi_\mm^\nn \sin^2\theta)}\right]
    \end{align}
\end{subequations}
\noindent where $k_0$ is the free-space wave-number. Under normal plane-wave incidence with $\theta=0^\circ$, the $\chi_\mm^\nn$ contribution vanishes, and the metasurface response depends only on tangential $\chi_\ee^\tt$, as expected. This response is compared with that obtained from HFSS and is shown in Fig.~\ref{fig:PracticalCell}(b). At non-normal incidence angles, if however, one chooses to ignore $\chi_\mm^\nn$, and only utilizes the tangential component to determine the angular scattering, it is clear from the comparisons in Fig.~\ref{fig:PracticalCell}(b) that the susceptibility model breaks down. As soon as the normal component is introduced, the angular scattering of the metasurface is perfectly modeled across the entire frequency range for all angles of incidences, as shown in Fig.~\ref{fig:PracticalCell}(c). Thus, this resonator based unit cell serves as a good example to illustrate that the normal surface susceptibilities play an indispensable role in accurate determination of angular scattering properties of metasurfaces. 

This example furthermore illustrates the usefulness of the surface susceptibility model. The practical structure of Fig.~\ref{fig:PracticalCell}(a) while being electrically thin, still has volumetric features, and two air-dielectric interfaces. The equivalent surface susceptibilities and the corresponding transmission and reflection functions of \eqref{eq:AnaTR}, on the other hand, represent a zero-thickness sheet model separating the two scattering regions by a single interface (or more precisely a spatial discontinuity). The determination of field scattering described using surface susceptibilities is thus expected to be computationally efficient compared to brute-force full-wave simulations of the volumetric structure, for specified incidence fields.

\begin{figure*}[htbp]
    \centering
   	 \begin{overpic}[width=1\linewidth,grid=false,trim={0cm 0cm 0cm 0cm},clip]   {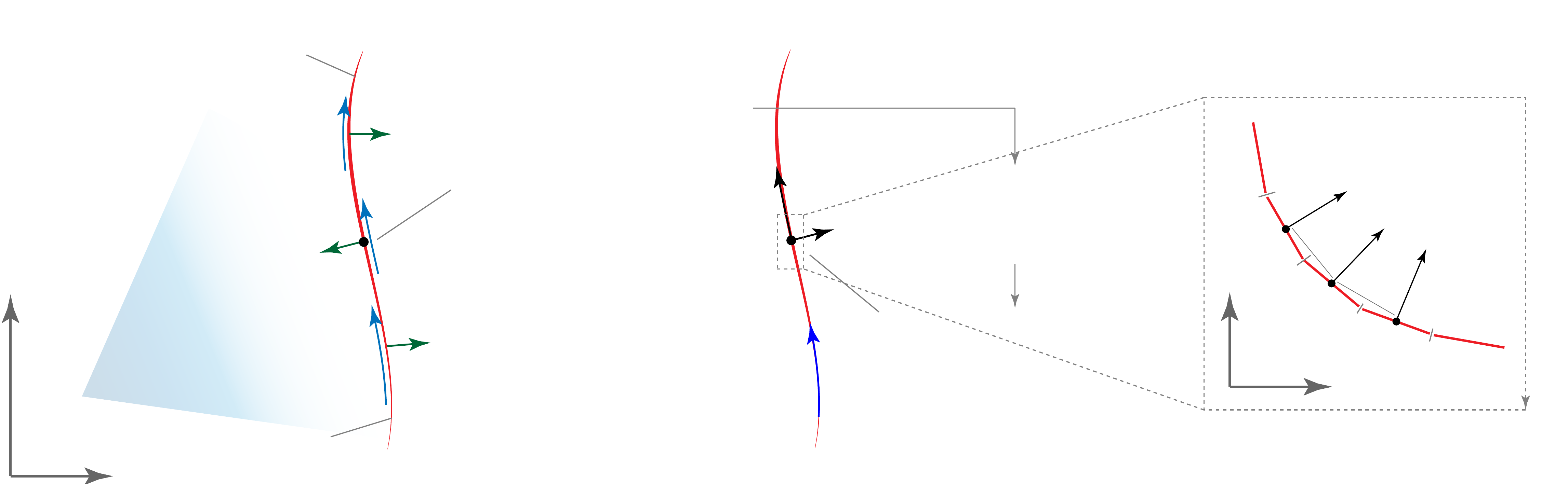}
        \put(9,  27){\htext{\scriptsize \shortstack{Zero-Thickness Metasurface Model \\ $\bar{\bar{\chi}}_\text{ee}(\r)$,~$\bar{\bar{\chi}}_\text{mm}(\r)$,~$\bar{\bar{\chi}}_\text{em}(\r)$,~$\bar{\bar{\chi}}_\text{me}(\r)$ }}}
        \put(14,  8){\htext{\scriptsize \shortstack{Specified Incident \\ Fields, $\psi^\text{inc.}(\r)$}}}
        \put(19,  2){\htext{\scriptsize $\delta=0$}}
         \put(16,  16){\htext{\scriptsize $\epsilon,~\mu$}}
         \put(34,  8){\htext{\scriptsize \shortstack{\color{ao}\textbf{Normal} Polarization \\ $\{P_n,~M_n\}$}}}
         \put(14,  21){\htext{\scriptsize \shortstack{\color{cobalt}\textbf{Tangential} Polarization \\ $\{P_t,~M_t\}$}}}
         \put(0.75,  12.5){\htext{\scriptsize $y$}}
         \put(8,  0.25){\htext{\scriptsize $x$}}
         \put(27,20){\scriptsize \shortstack{\textbf{Global} Surface Susceptibilities, $\bar{\bar{X}}$\\  $\begin{bmatrix} \chi_\text{ee}^{xx}&\chi_\text{ee}^{xy} & \chi_\text{ee}^{xz}\\\chi_\text{ee}^{yx}&\chi_\text{ee}^{yy}&\chi_\text{ee}^{yz}\\\chi_\text{ee}^{zx}&\chi_\text{ee}^{zy} &\chi_\text{ee}^{zz}\end{bmatrix}$}}
         \put(49,  21){\htext{\scriptsize $\hat{t}$}}
         \put(54,  16){\htext{\scriptsize $\hat{n}$}}
         \put(57, 6){\scriptsize \shortstack{\textbf{Local} Surface Susceptibilities, $\bar{\bar{\mathcal{X}}}$ \\ $\begin{bmatrix} \chi_\text{ee}^{nn}&\chi_\text{ee}^{nt} & \chi_\text{ee}^{nz}\\\chi_\text{ee}^{tx}&\chi_\text{ee}^{tt}&\chi_\text{ee}^{tz}\\\chi_\text{ee}^{zn}&\chi_\text{ee}^{zt} &\chi_\text{ee}^{zz}\end{bmatrix}$}}
         \put(65,  16.5){\htext{\scriptsize \color{amber}\shortstack{$\boxed{\chil = \Q \; \chit \; \Q^\Tt}$ \\ Co-ordinate Transformation}}}
         \put(87,  26){\htext{\scriptsize \shortstack{\color{ao}\textbf{Spatial Gradient} $\nabla_T\{\cdot\}$ \\ of Normal Surface Polarizations}}}
          \put(46.5,  6){\htext{\scriptsize \shortstack{Surface Currents \\ $\{\J,~\K\}$}}}
          \put(79.5,  15){\htext{\scriptsize $(i+1)$}}
          \put(83.5,  11.5){\htext{\scriptsize $i$}}
          \put(88,  8.5){\htext{\scriptsize $(i-1)$}}
          \put(88,  19.5){\htext{\scriptsize $\hat{n}_{i+1}$}}
          \put(89.5,  16.5){\htext{\scriptsize $\hat{n}_i$}}
           \put(93,  15){\htext{\scriptsize $\hat{n}_{i-1}$}}
           \put(88,  13){\htext{\tiny $l_{i-1/2}$}}
           \put(85,  16){\htext{\tiny $l_{i+1/2}$}}
            \put(86,  6){\htext{\scriptsize $x$}}
            \put(78.5,  13){\htext{\scriptsize $y$}}
                  \end{overpic}
        	\caption{The problem of field scattering from a zero-thickness metasurface sheet described in terms of dipolar tensorial surface susceptibilities, $\bar{\bar{X}}$, in the global co-ordinate system $\rh = (\xh,\yh,\zh)$, consisting of both tangential and normal surface polarizations. The susceptibilities are transformed into a surface susceptibilities, $\bar{\bar{\mathcal{X}}}$, in the local co-ordinate system $\lh = (\nh,\th,\zh)$ using the transformation matrix $\Q$. The zoomed region shows the details of surface showing three neighbouring segments for computing spatial gradient of the normal polarizations.} \label{fig:Problem}
\end{figure*}
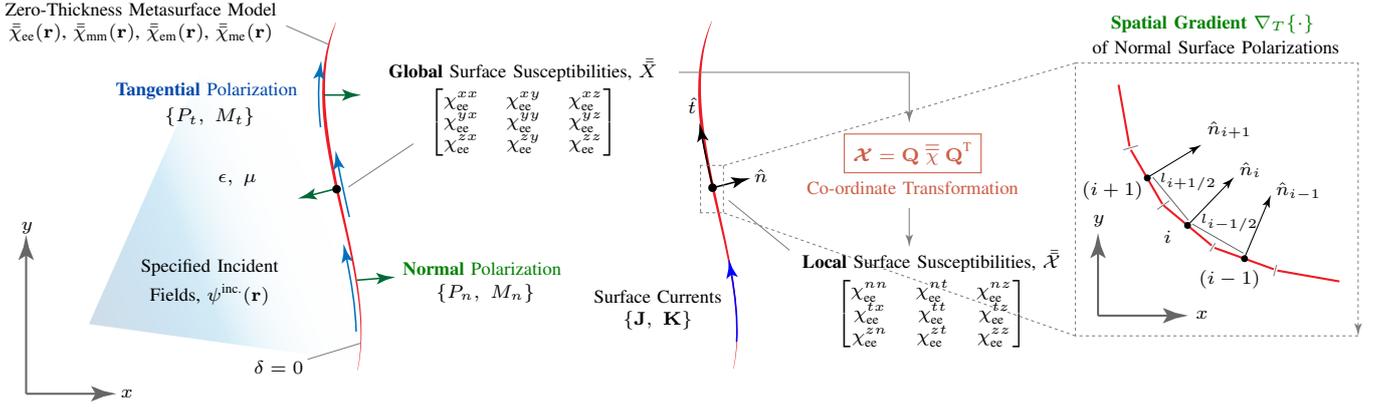

This problem is summarized in Fig.~\ref{fig:Problem}. Our objective now is to compute the scattered fields in the transmission and reflection regions of one or more  metasurfaces which are described in terms of their equivalent tensorial surface susceptibilities (with \emph{both} tangential and normal components) when excited with an arbitrary incident field, by solving the GSTCs embedded inside a bulk medium. It will be generally assumed next that for a practical metasurface of interest, the tensorial surface susceptibilities are already known or have already been obtained using standard extraction procedures using full-wave simulations or other experimental means \cite{GenBCEM}. 
% \vt{Comment: this objective discusses a single surface, and seems to contradict the problem statement at the beginning of the next section which discusses multiple surfaces. \tjs{fixed?}}

\subsection{Integral Equation (IE) based Field Propagation}

The most general statement of the problem of interest is the modeling of scattering of an EM field off one or more surfaces (media interfaces, metasurfaces, object surfaces etc.) -- including the coupling between surfaces. Solving the scattering problem is essentially the need to find a self-consistent solution of the excitation, the propagation of the fields and the interface conditions present for surfaces within the region of interest. If the surfaces present are uncharged, then these interface conditions relate the tangential components of the $\E$ and $\H$ fields across the interface or the surface. For electrically large objects placed in regions of uniform material parameters (electric permittivity $\epsilon$ and permeability $\mu$) an appropriate formulation of Maxwell's equations to determine the propagation is an Integral Equation (IE) approach. The equations describing the tangential field relationships at various interfaces are specific to that type of surface.

The EM fields radiated into free-space from electric and magnetic current sources, $\{\J,~\K\}$, can be generally expressed using an IE formulation as \cite{chew2009integral, Method_Moments}:
\begin{subequations}\label{Eq:FieldProp}
\begin{align}
    \E^\st(\r) &= -j\omega\mu(\L \J)(\r,\r') - (\R \K)(\r,\r')\\
    \H^\st(\r) &= -j\omega\epsilon(\L \K)(\r,\r') + (\R \J)(\r,\r'),
\end{align}
\end{subequations}
with $\r$ being the point of interest, $\r'$ the position of the source current; and  $\E^\st$ and $\H^\st$
the radiated (scattered) fields from the surface.\footnote{We will denote scattered or radiated fields due to the surface currents by the superscript $\st$ and total fields which include both scattered and incident fields  by a lack of superscript. Hence, generally $\E = \E^\st + \E^\text{i}$ where $\E^\text{i}$ is the incident field, for example.}  The field operators are given by:
\begin{align*}
    (\L \C)(\r, \r') &= \int_{\ell}[1+\frac{1}{k^2}\nabla\nabla\cdotp] [G(\r,\r')\C(\r')] \,d\r'\\
    (\R \C)(\r, \r') &= \int_{\ell}\nabla \times [G(\r,\r')\C(\r')] \,d\r'
\end{align*}
with $\C \in \{\J,\K\}$. $G(\r,\r')$ represents the Green's function which, for a 2D case, is given by the Hankel function of the \nth{2} kind, 
\begin{align*}
G(\r, \r')=    H_0^{(2)}(\r,\r') = J_0(\r,\r') - j Y_0(\r,\r'),
\end{align*}
where $J_0$ and $Y_0$ are the Bessel functions of the \nth{1} and \nth{2} kind and the function represents outwardly propagating radial waves. For numerical computation, these equations are discretized by segmenting the surfaces following Boundary Element Method (BEM) techniques and assembled in a matrix form for further processing, as will be shown later in Sec.~\ref{sec:bem}~\cite{chew2009integral,Method_Moments,stewart2019scattering}).

The above description is for the simulation of a set of surfaces in an infinite region of space. This is implicit in the formulation and determined by the choice of the Green's function (in this case a Hankel function of the 2$^\text{nd}$ kind). However, it can easily be extended to a \emph{periodic surface} by the use of a \emph{periodic Green's function}. The simplest approach to this is to simply tile the single period in the direction of periodicity. For example, if the problem is periodic along $y$ with a period length of $\Delta_y$ we have:
\begin{align}\label{Eq:PerGreen}
    G(\r, \r') &= \sum_{i = -\infty}^\infty H_0^{(2)}(\r,\r' + i \Delta_y)  \approx \sum_{i = -M}^M H_0^{(2)}(\r,\r' + i \Delta_y),
\end{align}
where the sum is truncated at $M$ after a sufficiently large number of terms to achieve convergence. This approach does not increase the size of the system matrix to be solved, but does require a large number of Hankel functions to be calculated to form the matrices. There have been attempts to speed up the convergence of this sum\cite{linton1998green}, however, for a 2D problem of moderate size, as we shall see, the problem is manageable when handled in a straight-forward way. It should be noted that the incident field contributions also have to be assumed periodic with an appropriate phase shift introduced for the tiled regions. 

\subsection{Surfaces in Local Co-ordinates}

If the region has no volumetric sources then $\J$ and $\K$ in \eqref{Eq:FieldProp} are only present on the defined interfaces and surfaces in the simulation region. These surface currents can be physically present or fictitious (a numerical construct) depending on the structure of the problem. For example, on a perfectly conducting electrical (PEC) surface the induced currents created by an incident field can be interpreted as physical electrical currents. However, for an interface between two dielectric regions with an incident field assumed to be present on a single side of the surface, the currents (both electric and magnetic) will be fictitious, but will produce the correct scattered fields in the reflection and transmission regions. 

For each surface (open or closed) present in the simulation domain, it is needed to formulate the surface field relationships that relate the total tangential fields on the two sides of the surface. Due to this requirement, it is convenient to express the surface characteristics in terms of a \emph{local coordinate system referenced to a surface normal}. For simplicity, we will assume wave propagation is 2D with all surfaces lying in the  $x$-$y$ plane and uniform along $z$ as shown in Fig.~\ref{fig:Problem}. For each surface point, we define a normal which lies in the $x$-$y$ plane $\nh = [n_x,n_y,0]$. An arbitrary vector $ \V = [V_x,V_y,V_z]$ present at the surface can be decomposed into a number of components using a local coordinate system referenced to that normal $\nh$, i.e. a tangential component defined by $\zh$ (as the surface extends in that direction), and a tangential direction $\th = - \nh \times \zh =[-n_y,n_x,0]$ which lies on the $x$-$y$ plane (see Fig.~\ref{fig:Problem}). We can, therefore, define a total tangential vector as,
\begin{align*}
    \V_\Tt = \bbmatrix V_t \\ V_z \ebmatrix,
\end{align*}
where $V_{t}$ is the tangential component of $\V$ in the $x$-$y$ plane and $V_z$ is the tangential component in the $\zh$ direction. We also define $V_n = \nh \cdot \V$ as the normal component of $\V$. We can reform the original vector,
\begin{align*}
    \V(\rh) = V_x \xh + V_y \yh + V_z \zh
\end{align*}
as,
\begin{align*}
    \Vl = V_{t} \th + V_n \nh + V_z \zh
\end{align*}
where the use of the calligraphic font indicates a quantity formulated in the local coordinate system. 

The new coordinate system is a rotation of the global coordinate system $\rh = (\xh,\yh,\zh)$ to a new local orthogonal system $\lh = (\nh,\th,\zh)$ where the {\it transformation} matrix $\Q$ rotates the vector through an angle $\theta$ in the $x$-$y$ plane, and is given by
\begin{align}\label{Eq:QMatrix}
    \Q = \bbmatrix \cos\theta & \sin\theta & 0\\ -\sin\theta & \cos\theta &0\\ 0 & 0 & 1\ebmatrix
    = \bbmatrix n_x & n_y & 0\\ -n_y & n_x &  0\\ 0 & 0 & 1\ebmatrix 
    \end{align}
so that
\begin{align*}
    \quad \Vl(\lh) = \Q \V(\rh).
\end{align*}
To extract the total tangent and normal vectors, we define the matrix operators $\Nl_\Tt$ and $\Nl_\ntt$:
\begin{align*}
    \Vl_\Tt &=  \bbmatrix V_{t} \\ V_z \ebmatrix = \bbmatrix 0 & 1 & 0\\ 0 & 0 & 1 \ebmatrix \bbmatrix V_n\\ V_{t} \\ V_z \ebmatrix
         = \Nl_\Tt \Vl \\
    V_n &= \bbmatrix 1 & 0 & 0 \ebmatrix \bbmatrix V_n\\ V_{t} \\ V_z \ebmatrix
         = \Nl_\ntt \Vl\\
\end{align*}
As we shall see, this coordinate system is useful to formulate the GSTCs, as it is invariant to the orientation of the surface and hence the surface susceptibilities will be as well. 

%\begin{figure}[b]
%    \centering
%    \begin{subfigure}[htbp]{0.5\columnwidth}
%        \includegraphics[width=0.9\linewidth]{Figures/Local2D.pdf}\caption{}
%	 \end{subfigure}\hfill
%%
% 	\begin{subfigure}[htbp]{0.5\columnwidth}
%  	 \includegraphics[width=1\linewidth]{Figures/2DSurf3points.pdf}\caption{}
%    \end{subfigure}
%        	\caption{The local and the global co-ordinate system. a) 3D Vector ($\V$) decomposition  into tangential ($V_{t}$ and $V_z$) and normal ($V_n$) components.  b) Detail of surface showing three neighboring segments for computing spatial gradient of the normal polarizations.} \label{fig:metas}
%\end{figure}

\section{GSTCs in a Local Co-ordinate System}\label{Sec:Form}

\subsection{Metasurface Description}\label{Sec:II^-B}

Lets us look next at how the relationship between the fields across a given surface/interface (and more generally a metasurface) may be described in its local coordinate system. The simplest surface is a \emph{Perfect Electric Conductor (PEC)} where the tangential electric field on both sides is equal to zero. This can be described by,
\begin{align} \label{eq:PEC}
\bbmatrix
      \Nl_{T} & \0 & \0  & \0\\
      \0 & \0 & \Nl_{T}  & \0\\
\ebmatrix 
\bbmatrix \El^+\\\Hl^+\\\El^-\\\Hl^- \ebmatrix
  &= \bbmatrix \0 \\ \0 \ebmatrix
\end{align}
where $\{\cdot\}^+$ and $\{\cdot\}^-$ indicate the positive and negative surface sides, respectively with respect to the surface normal $\nh$. The matrix operator $\Nl_\Tt$ performs the operation of extracting the two tangential fields at the surface (one in the $x-y$ plane and the other with respect to $\zh$) obtaining $\El_\Tt$ from $\El$ for example. For a \emph{dielectric interface}, the tangential E- and H-fields on both sides are equal and thus we have,
\begin{align}\label{eq:Die}
\left[ \begin{array}{cccc}
     \Nl_{T} & \0 & -\Nl_{T}  & \0\\
     \0 &\Nl_{T} &\0 & -\Nl_{T}
\end{array}\right] \left[ \begin{array}{c}\El^+\\\Hl^+\\\El^-\\\Hl^-\end{array}\right]
 &= \left[ \begin{array}{cccc} \0\\\0
\end{array}\right]
\end{align}
A much more general surface formulation is using the GSTCs, which we shall now address using the local coordinate frame work. An electromagnetic metasurface as remarked in the introduction, can be rigorously described using a zero thickness sheet model, using Generalized Sheet Transition Conditions (GSTCs) with four sets of tensorial surface susceptibilities $\chit_{\alpha\beta}(\r_m, \omega)$ with $\alpha\beta \in \{ \text{ee, mm, em me} \}$. This formulation captures the general wave transformation capability of physical EM metasurfaces by expressing them as mathematical space discontinuities of zero thickness \cite{IdemenDiscont,GSTC_Holloway, KuesterGSTC}. 
The  GSTCs relate the tangential EM fields around the metasurface to the tangential and normal surface polarization response, rigorously modeling the EM interaction with the metasurface capturing the field transformation capabilities via 36 variables inside the susceptibility tensors. 
% \vt{[Note: the previous 2 sentences seem a bit repetitive to me considering they are already in the introduction. Perhaps they can be omitted.]} 
The  GSTC formulation in the global coordinate system $\rh$, accounting for tangential and normal polarizations is given by \cite{KuesterGSTC}:
\begin{subequations} \label{eq:GSTC}
\begin{align}
	\Delta \E &= j\omega (\Nh \times  \M) - \nabla_{T}\left(\frac{P_n}{\epsilon}\right)\\
	\Delta \H  &= -j\omega (\Nh \times\P) - \nabla_{T}\left(\frac{M_n}{\mu}\right)
\end{align}
\end{subequations}
where $\Delta \psi_\Tt = \psi^+ - \psi^-$,
% \vt{\sout{, and $\psi_\av = \{\psi^+ + \psi^-\}/2$,}}
are expressed in terms of total fields just before and after the metasurface, and $\{\P,~\M\}$ are the electric and magnetic polarizations, respectively.

Restricting ourselves to the tangential components, as they are what needs to be matched across the interface, the GSTCs can also be expressed in the local coordinate system $\lh$ as:
\begin{subequations} \label{eq:GSTSs}
\begin{align}
	\Delta \El_\Tt &= j\omega (\Nh \times  \Ml)_\Tt - \nabla_{T}\left(\frac{\Pln}{\epsilon}\right)\\
	\Delta \Hl_\Tt  &= -j\omega (\Nh \times\Pl)_\Tt - \nabla_{T}\left(\frac{\Mln}{\mu}\right)
\end{align}
\end{subequations}
with $[\cdot]_\Tt$ signifying the tangential components only. Defining the polarizations (also in the local coordinate system) we have,
\begin{subequations}
\begin{align*}
    \Pl &= \epsilon \chil_\ee \El_\av + \chil_\emm \sqrt{\mu \epsilon}\; \Hl_\av\\
    \Ml &= \mu \chil_\mm \Hl_\av + \chil_\me \sqrt{\mu \epsilon}\; \El_\av
\end{align*}\label{eq:ConstitutiveRelations}
\end{subequations}
with $\psi_\av = \{\psi^+ + \psi^-\}/2$ and where we define the \emph{surface susceptibility dyadics} as
\begin{align*}
    \chil_{\alpha\beta}  = \bbmatrix \chi^\nn_{\alpha\beta}&\chi^\nt_{\alpha\beta}&\chi^\nz_{\alpha\beta}\\
                                 \chi^\tn_{\alpha\beta}&\chi^\tt_{\alpha\beta}&\chi^\tz_{\alpha\beta}\\
                                 \chi^\zn_{\alpha\beta}&\chi^\zt_{\alpha\beta}&\chi^\zz_{\alpha\beta}\ebmatrix,
\end{align*}
with $\alpha\beta \in \{\text{ee,~mm,~em,~me}\}$. Naturally, the surface susceptibility dyadics above are also defined in the local coordinate system $\lh$. 

%\begin{figure}[htbp]
%    \centering
%    \begin{subfigure}[htbp]{0.5\columnwidth}
%      \includegraphics[width=1\textwidth]{Figures/MetaStr.pdf}\caption{}
%	 \end{subfigure}\hfill
%%
% 	\begin{subfigure}[htbp]{0.5\columnwidth}
%  	 \includegraphics[width=1\linewidth]{Figures/2DSurf3points.pdf}\caption{}
%    \end{subfigure}
%        	\caption{Curvilinear metasurface showing invariant $\chi$'s with respect to the normal. Detail of surface showing three neighboring segments.} \label{fig:metas}
%\end{figure}

Typically, the surface susceptibilities of metasurface unit cells are conveniently extracted using a global co-ordinate systems $\rh$, with periodic boundaries and using a specific set of incident plane waves. These susceptibilities are naturally with reference to the surface normal of the unit cell with respect to $\rh$. If the same unit cell is now used to construct a \emph{curvilinear metasurface} or if the metasurface has a spatial orientation different from that used in the susceptibility extraction process, these global surface susceptibilities $\chit$ must be reoriented to the local co-ordinates $\lh$. The relationship of these local susceptibilities to the global equivalents is provided by the transformation matrix at each point on the surface, as illustrated in Fig.~\ref{fig:Problem}:
\begin{align*}
    \chil = \Q \; \chit \; \Q^\Tt.
\end{align*}
This ability to conveniently model an arbitrary shaped curvilinear metasurface is a key advantage of utilizing a local co-ordinate system as opposed to a global one.

%A key advantage of this formulation in the local coordinate form is that for a curvilinear metasurface the susceptibilities are invariant (at least for a uniform surface where all the unit cells are identical). This is illustrated in Fig.~\ref{fig:metas}. If it is wished to transform from susceptibilities ($\chit$) expressed in the global coordinate system ($\rh$) to ones ($\chil$) formulated for the local coordinates ($\lh$) the transformation matrix can be used:
%\begin{align*}
%    \chil = \Q \; \chit \; \Q^\Tt
%\end{align*}
%
\begin{figure*}[htbp] \hrulefill \par
\begin{subequations}\label{Eq:GSTC_LCS}
	\begin{equation}
	\Delta (\Nl_\Tt \El) = j\omega \mu \Nl_\Tt \Rlx \chil_\mm \Hl_\av +  j\omega \sqrt{\mu \epsilon}\; \Nl_\Tt \Rlx \chil_\me \El_\av - \nabla_{T}\left(\frac{\Nl_\ntt \epsilon \chil_\ee \El_\av + \Nl_\ntt \chil_\emm \sqrt{\mu \epsilon}\; \Hl_\av}{\epsilon}\right)
\end{equation} 
\begin{equation}
	\Delta (\Nl_\Tt \Hl)  = - j\omega \epsilon \Nl_\Tt \Rlx\chil_\ee \El_\av - j\omega   \sqrt{\mu \epsilon}\; \Nl_\Tt \Rlx \chil_\emm \Hl_\av - \nabla_{T}\left(\frac{\Nl_\ntt \mu \chil_\mm \Hl_\av + \Nl_\ntt \chil_\me \sqrt{\mu \epsilon}\; \El_\av}{\mu}\right) 
	\end{equation}
\end{subequations}   \hrulefill \par
\end{figure*}

\begin{figure*}[t] \hrulefill \par

\begin{subequations}\label{Eq:GSTC_GL}
	\begin{equation}
	\Delta (\Nl_\Tt \Q \E) = j\omega \mu \Nl_\Tt \Rlx \chil_\mm \Q \H_\av +  j\omega \sqrt{\mu \epsilon}\; \Nl_\Tt \Rlx \chil_\me \Q \E_\av - \nabla_{T}\left(\frac{\Nl_\ntt \epsilon \chil_\ee \Q \E_\av + \Nl_\ntt \chil_\emm \sqrt{\mu \epsilon}\; \Q \H_\av}{\epsilon}\right)
\end{equation} 
\begin{equation}
	\Delta (\Nl_\Tt \Q \H)  = - j\omega \epsilon \Nl_\Tt \Rlx\chil_\ee \Q \E_\av - j\omega   \sqrt{\mu \epsilon}\; \Nl_\Tt \Rlx \chil_\emm \Q \H_\av - \nabla_{T}\left(\frac{\Nl_\ntt \mu \chil_\mm \Q \H_\av + \Nl_\ntt \chil_\me \sqrt{\mu \epsilon}\; \Q \E_\av}{\mu}\right)
	\end{equation}
\end{subequations}   \hrulefill \par
\end{figure*}

Next, we wish to express \eqref{eq:GSTSs} using matrix operators suitable for incorporation into the BEM framework to implement IE-GSTCs. To reform \eqref{eq:GSTSs}, we need to express $(\nhl \times \Pl)_\Tt$ and $(\nhl \times \Ml)_\Tt$ as matrix operators. The cross-product operator is simple in the local coordinate system as the normal is the first orthogonal direction and we have $\nhl = [1 \; 0 \; 0]$ which results in,
\begin{align*}
    \nhl \times \Vl 
    % &= \left| \begin{array}{ccc}i & j & k\\ \hline 1 & 0 &0\\ V_n & V_{t} & V_z \end{array}\right|\\  
    &= \bbmatrix 0 & 0 & 0\\ 0 & 0 & -1\\ 0 & 1 & 0 \ebmatrix \bbmatrix V_n \\ V_{t} \\ V_z \ebmatrix =  \Rlx \Vl
\end{align*}
The two tangential terms of interest in the right-hand side of \eqref{eq:GSTSs} therefore become,
\begin{align*}
    (\nhl \times \Pl)_\Tt &=  \epsilon\; \Nl_\Tt \Rlx\chil_\ee \El_\av +  \sqrt{\mu \epsilon}\; \Nl_\Tt \Rlx \chil_\emm \Hl_\av\\
    (\nhl \times \Ml)_\Tt &=  \mu\; \Nl_\Tt \Rlx \chil_\mm \Hl_\av +  \sqrt{\mu \epsilon}\; \Nl_\Tt \Rlx \chil_\me \El_\av
\end{align*}
where all vectors and tensors are in the local coordinate system. We can also extract the normal components of the surface polarizations using the operator $\Nl_\ntt$ as,
\begin{align*}
    \mathcal{P}_n &= \Nl_\ntt \Pl = \Nl_\ntt \epsilon \chil_\ee \El_\av + \Nl_\ntt \chil_\emm \sqrt{\mu \epsilon}\; \Hl_\av\\
    \mathcal{M}_n &= \Nl_\ntt \Ml = \Nl_\ntt \mu \chil_\mm \Hl_\av + \Nl_\ntt \chil_\me \sqrt{\mu \epsilon}\; \El_\av
\end{align*}

We can now reform \eqref{eq:GSTSs} completely in terms of the local coordinate system using matrix operators resulting in \eqref{Eq:GSTC_LCS}. However, the field propagation equations of Sec.~II-B, must be expressed in the global coordinate system $\r$ and in order to link the GSTCs of \eqref{Eq:GSTC_LCS}, we need to express it in terms of the fields in the global coordinate system. To achieve this, we can use \mbox{$\Vl=\Q \V$} to convert the fields from local to global coordinates, which transforms 
\eqref{Eq:GSTC_LCS} into \eqref{Eq:GSTC_GL}.

If we further define the following quantities, 
\begin{align*}
    \XlT_\mm &= j\omega \mu \Nl_\Tt \Rlx \chil_\mm \Q\\
    \XlT_\me &= j\omega \sqrt{\mu\epsilon}\Nl_\Tt \Rlx \chil_\me \Q\\
    \XlT_\ee &= - j\omega \epsilon \Nl_\Tt \Rlx \chil_\ee \Q\\
    \XlT_\emm &= -j\omega \sqrt{\mu\epsilon} \Nl_\Tt \Rlx \chil_\emm \Q \\
    \Xln_\ee &= \Nl_\ntt \chil_\ee \Q\\
    \Xln_\me &= \sqrt{\mu/\epsilon}\; \Nl_\ntt \chil_\me \Q\\
    \Xln_\mm &= \Nl_\ntt \chil_\mm \Q\\
    \Xln_\emm &= \sqrt{\epsilon/\mu}\; \Nl_\ntt \chil_\emm \Q\\
    \Nu_\Tt &= \Nl_\Tt \Ql
\end{align*}
\eqref{Eq:GSTC_GL} can be expressed more compactly as:
\begin{subequations}\label{eq:gstc_loc}
\begin{align}
	\Delta (\Nu_\Tt \E) &= \XlT_\mm \H_\av +  \XlT_\me \E_\av - \notag\\ 
	&\quad \nabla_\Tt\left(\Xln_\ee \E_\av + \Xln_\emm\H_\av\right)\\
	\Delta (\Nu_\Tt \H)  &= \XlT_\ee \E_\av + \XlT_\emm \H_\av - \notag\\ 
	&\quad \nabla_\Tt\left(\Xln_\mm\H_\av + \Xln_\me\E_\av\right)
\end{align}
\end{subequations}
This formulation has the fields in the global coordinate system $\r$ but the susceptibilities in the local coordinate frame $\lh$. The final operation that we need to represent as a matrix operator is the gradient operator $\nabla_{\Tt}$ which operates on the normal component of the surface polarizations. 

\subsection{Gradient of the Normal Susceptibilities}

In above formulation we have only needed to deal with a single point on the surface to express the GSTCs. This was due to the incorporation of the analytical gradient operator $\nabla_\Tt$. However, to proceed to a complete matrix formulation, we will need to discretize the metasurface and be able to compute the spatial gradient along the surface using neighboring points, as illustrated in the zoomed region of Fig.~\ref{fig:Problem}. We shall use a central difference methodology to evaluate the gradient using a segment on the surface and its nearest neighbours. 

The operator $\nabla_\Tt$ is expressed as two components in the local coordinate system and given that the fields are invariant along $\zh$, we have,
\begin{align*}
    \nabla_\Tt = \bbmatrix \nabla_{t}\\ \nabla_z \ebmatrix = \bbmatrix \nabla_{t}\\ 0 \ebmatrix
\end{align*}
The terms we want to formulate have the form of,
\begin{align*}
      \nabla_\Tt (\Xln \V) = \nabla_\Tt (\Nl_\ntt \chil \Q \V) = \nabla_\Tt F_n
\end{align*}
where $F_n = \Nl_\ntt \chil \Q \V$ is the projection of the polarization along $\nh$. As we need to take a gradient of $F_n$ along the surface in the tangential directions, we define collections of sets of 3 vectors such as.
\begin{align*}
    \Vf = \bbmatrix \V_{i-1} \\ \V_{i}\\ \V_{i+1} \ebmatrix
\end{align*}
where the gothic font is used to indicate collections of three sequential scalars or vectors; or operators that act on such collections. 
Collecting the normal polarization contributions into a triplet we obtain,
\begin{align*}
\Ff = \bbmatrix F_{n,i-1} \\ F_{n,i}\\F_{n,i+1}\ebmatrix = \bbmatrix \Xln_{i-1} & \0 & \0\\ \0 & \Xln_{i} & \0\\ \0 & \0& \Xln_{i+1}\\\ebmatrix 
\bbmatrix \V_{i-1} \\ \V_{i}\\ \V_{i+1}\ebmatrix = \Xf^n \Vf
\end{align*}
The gradient in the $x$-$y$ plane in the direction of the tangent $\th$ is given using a centered finite difference approach by,
\begin{align*}
    \nabla_{t} F_{n,i} &= \frac{1}{2} \left[\frac{F_{n,i+1} - F_{n,i}}{l_{i+1/2}} +  \frac{F_{n,i} - F_{n,i-1}}{l_{i-1/2}} \right]\\
    &= \frac{1}{2} \bbmatrix -l_{i-1/2}^{-1} & l_{i-1/2}^{-1} - l_{i+1/2}^{-1} & l_{i+1/2}^{-1} \ebmatrix \bbmatrix F_{n,i-1} \\F_{n,i}\\F_{n,i+1}\ebmatrix
\end{align*}
As the gradient of all quantities (2D assumption) in the $\zh$ direction is zero we define,    
\begin{align*}
    \Lf_i&= \frac{1}{2} \bbmatrix -l_{i-1/2}^{-1} & (l_{i-1/2}^{-1} - l_{i+1/2}^{-1}) & l_{i+1/2}^{-1}\\0&0&0 \ebmatrix 
    % \bbmatrix F_{n,i-1}\\F_{n,i}\\F_{n,i+1} \ebmatrix \\
\end{align*}
and have\footnote{Care needs to be taken with this operator at the ends of the surface. For a periodic surface the operator should be periodic. For a finite freestanding surface the operator should make an implicit assumption that there is a transparent dielectric extension.}, 

\begin{align*}
    \nabla_\Tt \bbmatrix F_{n,i}\\ 0 \ebmatrix &= \Lf_i \Ff_i = \Lf_i \Xf_i^n \Vf_i = \Lf_i^{\Xf} \Vf_i
\end{align*}
The GSTCs of \eqref{eq:gstc_loc} finally becomes in a full matrix operator form as,
\begin{subequations}\label{Eq:GSTC_Mat}
\begin{align*}
	\Delta (\Nu_\Tt \E) &= \XlT_\mm \H_\av +  \XlT_\me \E_\av+ \Lf^{\Xf}_\ee \Ef_\av + \Lf^{\Xf}_\emm\Hf_\av\\
 	\Delta (\Nu_\Tt \H)  &= \XlT_\ee \E_\av + \XlT_\emm \H_\av + \Lf^{\Xf}_\mm\Hf_\av + \Lf^{\Xf}_\me\Ef_\av
\end{align*}
\end{subequations}

We can further manipulate it by defining a surface field vector for the two sides of the metasurfaces, which collects fields for all three points, 
\begin{align*}
    \Sf_\Ft = [\Ef^{+} \; \Hf^{+} \; \Ef^{-} \; \Hf^{-}]
\end{align*}
and an appropriate tangent operator
\begin{align*}
    \Nf_\TQt = [\0 \; \Nu_\Tt \; \0 ]
\end{align*}
This finally allows us to express the GSTCs of \eqref{Eq:GSTC_Mat} in a very compact form, using \emph{susceptibilities expressed in a local coordinate frame}, the \emph{fields expressed in the global coordinate frame}, and \emph{matrix operators} operating on the electric and magnetic surface polarizations (tangential and normal) as:
\begin{align}\label{eq:GSTCcom}
    \Df_\text{TF} \Sf_\Ft = \Gf_\text{TF} \Sf_\Ft,
\end{align}
with the following defined matrix operators:
\begin{align*}
\Df_\text{TF} &= \bbmatrix
     \Nf_\TQt & \0 & -\Nf_\TQt  & \0\\
     \0 &\Nf_\TQt &\0 & -\Nf_\TQt
\ebmatrix \\
\Gf_\text{TF} &=  \bbmatrix
     \Gf_\me  & \Gf_\mm & \Gf_\me^{-} & \Gf_\mm\\
     \Gf_\ee  & \Gf_\emm & \Gf_\ee^{-} & \Gf_\emm\\
\ebmatrix
\end{align*}
and,
\begin{subequations}%\label{eq:Gs}
\begin{align*}
    \Gf_\ee &= \frac{1}{2}\left(\bbmatrix \0 & \XlT_\ee& \0 \ebmatrix + \Lf^{\Xf}_\me\right)\\
    \Gf_\me &= \frac{1}{2}\left(\bbmatrix \0 & \XlT_\me& \0 \ebmatrix + \Lf^{\Xf}_\ee\right)\\
    \Gf_\mm &= \frac{1}{2}\left(\bbmatrix \0 & \XlT_\mm& \0 \ebmatrix + \Lf^{\Xf}_\emm\right)\\
    \Gf_\emm &= \frac{1}{2}\left(\bbmatrix \0 & \XlT_\emm& \0 \ebmatrix + \Lf^{\Xf}_\mm\right).
\end{align*}
\end{subequations}

\section{IE-GSTCs and BEM Implementation} \label{sec:bem}

The GSTCs presented above represent the general relationship between the fields across a metasurface, in terms of the surface polarizations $\P$ and $\M$, which in turn may be related to electric and magnetic currents $\J$ and $\K$ on the metasurface \cite{Smy_Close_ILL, smy2020IllOpen, stewart2019scattering}. The unknown scattered fields from the metasurface in response to a specified incident fields can thus be obtained as self-consistent solutions of IE of \eqref{Eq:FieldProp} and metasurface boundary conditions given by the GSTCs (\ref{eq:GSTCcom}). These set of fields equations together represents the IE-GSTC formulation of general metasurfaces, which now must be solved. A well known approach to solving EM scattering problems such as described by \eqref{Eq:FieldProp} is the Boundary Element Method (BEM) whereby the surfaces are spatially discretized so that the integral equations representing propagation become a set of algebraic equations. The appropriate interface equations for the tangential fields [e.g. \eqref{eq:PEC}, \eqref{eq:Die} and (\ref{eq:GSTCcom})] are then used to complement the propagation equations and a self-consistent solution can be obtained for a known incident field. 

 Each discrete element of the metasurface is characterized by a center position $\r_{\ptt,i} = [x_\ptt ,y_\ptt ,0]_i$, a length $\ell_i$ and a normal $\Nh_i$ which are collected into vectors such as $\rbb_\ptt = \bbmatrix \r_{\ptt,1} & \dots & \r_{\ptt,m} \ebmatrix$. The field quantities are stored in vectors of the same form (with $m$ surface elements):
\begin{align*}
    \Ev &= \bbmatrix \E_1 & \E_2 & \dots & \E_{m} \ebmatrix\\
    \Hv &= \bbmatrix \H_1 & \H_2 & \dots & \H_{m} \ebmatrix\\
    \Sv_\Ft &= \bbmatrix \Ev^- & \Hv^- &\Ev^+ & \Hv^+  \ebmatrix
\end{align*}
The fields are stored sequentially along the surface and can also be thought of as triplets of fields ($\Ef_i$ and $\Hf_i$) along the surface.  This will facilitate the use of the operators defined in Sec.~III, to calculate the gradient of the normal components of the polarization.

\subsection{Field Propagation}

The EM fields radiated from each discretized surface element due to electric and magnetic surface currents can be generally expressed as  discretization of \eqref{Eq:FieldProp}:\cite{stewart2019scattering,chew2009integral,Method_Moments,smy2020IllOpen}
\begin{align}\label{eq:BEMProp}
     \bbmatrix \Ev^\st(\rbb_\ptt) \\\Hv^\st(\rbb_\ptt) \ebmatrix & = 
    \bbmatrix - j\omega \mu \Lv(\rbb _\ptt, \rbb_\St)  & - \Rv(\rbb_\ptt, \rbb_\St)  \\ 
             \Rv(\rbb_\ptt, \rbb_\St)  & - j\omega \epsilon \Lv(\rbb_\ptt, \rbb_\St) \ebmatrix 
    \bbmatrix  \Jv_\St \\ \Kv_\St\ebmatrix  
%     \Fv_{p,S} & = \Pv_{S,p} \Cv_\St
\end{align}
where $\rbb_\ptt$ is a set of points at which the fields are desired, and $\rbb_\St$ are the points on the surfaces where the source currents are defined. The discretized field propagation matrices $\Lv$ and $\Rv$ are given by (for example),
\begin{align*}
\Lv(\rbb_\ptt, \rbb_\St) &= \bbmatrix 
\L_{1,1} & \L_{1,2} & \dots &\L_{1,m} \\
\L_{2,1} & \L_{2,2} & \dots &\L_{2,m} \\
\vdots & \vdots & \ddots & \vdots\\
\L_{n,1} & \L_{n,2} & \dots &\L_{n,m} \\
\ebmatrix
\end{align*}
with 
\begin{align*}
    \L_{i,j} %&= \L(\rb_{p,i},\rb_{S,j})   \\
    &= \int_{\delta\ell_j}[1+\frac{1}{k^2}\nabla_{\rb_\ptt}\nabla_{\rb_\ptt}\cdotp] [G(\r_{\ptt,i},\r_{\St,j})] \,d\r_{\St,j}
\end{align*}
and the complementary operator used to form  $\Rv$ is,
\begin{align*}
    \R_{i,j} %&= \R(\rb_{p,i},\rb_{S,j})   
     = \int_{\delta\ell_j}\nabla_{\rb_\ptt} \times [G(\r_{\ptt,i},\r_{\St,j})] \,d\r_{\St,j}
\end{align*}
If we set $\rbb_\ptt = \rbb_\St$ then \eqref{eq:BEMProp} can be used to determine the fields generated at the surface due to given excitation fields along with the unknown surface currents $\Jv_\St$ and $\Kv_\St$. However, in order to do this we need a discretized formulation for the tangential components of the fields at the surfaces. 

\subsection{Surface Formulation}

To describe a set of surfaces which are either PEC, dielectric interfaces or described by the GSTC we need versions of \eqref{eq:PEC}, \eqref{eq:Die} or \eqref{eq:gstc_loc} for a collection of surface elements forming a discretized surface. Using the PEC as a simple example we first note that \eqref{eq:PEC} needs to be reformulated in terms of the global coordinate system $\rb$ to comport with the propagation equations. We can use the transformation matrix $\Q$ of \eqref{Eq:QMatrix} to rewrite \eqref{eq:PEC} as, 
\begin{align*}
\bbmatrix
      \Nl_\Tt\Q & \0 & \0  & \0\\
      \0 & \0 & \Nl_\Tt\Q  & \0\\
\ebmatrix 
\bbmatrix \E^+\\\H^+\\\E^-\\\H^- \ebmatrix
  &= \bbmatrix \0 \\ \0 \ebmatrix
\end{align*}
Each matrix $\Nl_\Tt\Q$ is of the form,
\begin{align} \label{eq:NTQ}
    \Nl_\Tt\Q = \Nu_\TQt = 
    \bbmatrix N_\TQt^\text{tx} &  N_\TQt^\text{ty}  & 0 \\
              0 & 0   & N_\TQt^\nn\\
    \ebmatrix
\end{align}
We can then collect the terms into a matrix that extracts the tangential components of a field over the entire surface by placing the matrix operator $\Nu_\TQt$ along the diagonal,
\begin{align*}
    \Nv_\TQt  = \bbmatrix \Nu_{\text{TQ},1} & \dots & \0 \\
                         \0 &\ddots  &\0 \\
                         \0 & \0 & \Nu_{\text{TQ},m} \\
                \ebmatrix
\end{align*}
The surface conditions for PEC surface can then be enforced by,
\begin{align}\label{eq:PECSurf}
\bbmatrix
      \Nv_\TQt & \0 & \0  & \0\\
      \0 & \0 & \Nv_\TQt  & \0\\
\ebmatrix 
\bbmatrix \Ev^+\\\Hv^+\\\Ev^-\\\Hv^- \ebmatrix
  &= \bbmatrix \0 \\ \0 \ebmatrix \notag\\
  \Rightarrow \boxed{\Nv_\text{TS} \Sv_\Ft = \0}
\end{align}

\newcommand{\Pt}{\text{P}}
\newcommand{\Gt}{\text{G}}
\newcommand{\Ct}{\text{C}}
\newcommand{\TFt}{\text{TF}}
\newcommand{\GFt}{\text{GF}}
\newcommand{\PFt}{\text{PF}}
\newcommand{\CSt}{\text{CS}}

A similar relationship can be derived for a dielectric interface from \eqref{eq:Die}. We define a tangent field difference operator for a surface:
\begin{align*}
\Dv_\TFt = \bbmatrix
      \Nv_\TQt & \0 & -\Nv_\TQt  & \0\\
      \0 & \Nv_\TQt & \0 & -\Nv_\TQt 
\ebmatrix 
\end{align*}
so that the dielectric interface conditions for a surface are,
\begin{align}\label{eq:DieSurf}
   \boxed{ \Dv_\TFt\Sv_F = \0}
\end{align}

Finally, the transformation of the GSTC equation \eqref{eq:GSTCcom} results in the following matrix equation,
\begin{align*}
&\bbmatrix
     \Nv_\TQt & \0 & -\Nv_\TQt  & \0\\
     \0 &\Nv_\TQt &\0 & -\Nv_\TQt
\ebmatrix \bbmatrix \Ev^{+}\\\Hv^{+}\\\Ev^{-}\\\Hv^{-}\ebmatrix
 = \\
 &\bbmatrix
     \Gv_\me^{+}  & \Gv_\mm^{+} & \Gv_\me^{-} & \Gv_\mm^{-}\\
     \Gv_\ee^{+}  & \Gv_\emm^{+} & \Gv_\ee^{-} & \Gv_\emm^{-}
\ebmatrix
\bbmatrix \Ev^{+}\\\Hv^{+}\\\Ev^{-}\\\Hv^{-}\ebmatrix
\end{align*}
or 
\begin{align}\label{eq:GSTCSurf}
    \boxed{\Dv_\TFt \Sv_F = \Gv_\TFt \Sv_F}
\end{align}
where each of the sub-matrices forming $\Gv_\TFt$ are of the form,
\begin{align*}
    \Gv_{\alpha\beta}  = \bbmatrix \Gf_{\alpha\beta,1} & \dots & \0 \\
                         \0 &\ddots  &\0 \\
                         \0 & \0 & \Gf_{\alpha\beta,m} \\
    \ebmatrix,
\end{align*}
which is a diagonal matrix created from triplet operators.

\subsection{Full IE-GSTC Matrix Solution}

For a simulation region comprised of a set of discretized surfaces and an incident excitation, the propagation equations  \eqref{eq:BEMProp} and the appropriate surface equations can be coupled to produce a field scattering solution. It is convenient to define a propagation matrix $\Pv_{\St\St'}$ from the surface $\St$ to the surface $\St'$ along with a surface current vector $\Cv_\St$, 
\begin{align*}
    \Pv_{\St\St'} &= \bbmatrix
    - j\omega \mu \Lv(\rbb_{\St'}, \rbb_\St)  & - \Rv(\rbb_{\St'}, \rbb_\St) \\ 
    \Rv(\rbb_{\St'}, \rbb_\St)  & - j\omega \epsilon \Lv(\rbb_{\St'}, \rbb_\St) 
    \ebmatrix\\
    \Cv_\St &= \bbmatrix \Jv_\St & \Kv_\St \ebmatrix
\end{align*}
We need to constrain the currents to be tangential to the surface (i.e. no normal component). To do this we construct a matrix operator,
\begin{align*}
    \NN_\NCt = \bbmatrix \NN_\ntt & \0\\ \0 & \NN_\ntt \ebmatrix
\end{align*}
with 
\begin{align*}
    \Nv_{\ntt}  = \bbmatrix \Nu_{\ntt,1} & \dots & \0 \\
                         \0 &\ddots  &\0 \\
                         \0 & \0 & \Nu_{\ntt,m} \\
                \ebmatrix
\end{align*}
and impose the condition that the normal current flow is zero,
\begin{align}\label{Eq:NormC0}
    \NN_\NCt \Cv_\St = \bbmatrix \NN_\ntt & \0\\ \0 & \NN_\ntt \ebmatrix \bbmatrix \Jv_\St \\\Kv_\St \ebmatrix  = \0
\end{align}
As an example, for a region with two surfaces, consider a PEC surface ($P$) and the other a metasurface ($G$), a system matrix can be constructed by combining the propagation equations [\eqref{eq:BEMProp}] and the surface field relationships [\eqref{eq:PECSurf}, \eqref{eq:DieSurf} or \eqref{eq:GSTCSurf}], and further constraining the currents to flow only tangentially on the surfaces [\eqref{Eq:NormC0}], leading to a single matrix equation: 
\begin{align*}
    \bbmatrix 
        \Pv_{\Pt\Pt} & \Pv_{\Gt\Pt} & -\I & \0\\
        \Pv_{\Pt\Gt} & \Pv_{\Gt\Gt} & \0 & -\I\\
        \Nv_\text{CP} & \0 & \0 & \0\\
        \0 & \Nv_\text{CG} & \0 & \0\\
        \0 & \0 & \Nv_\text{TS}  & \0 \\
        \0 & \0 & \0 &\Dv_\TFt-\Gv_\TFt  \\
    \ebmatrix
    &
    \bbmatrix \Cv_\Pt \\\Cv_\Gt \\\Sv^\st_\PFt\\\Sv^\st_\GFt \ebmatrix
    = 
    \bbmatrix \0 \\ \0 \\ \0 \\ \0 \\-\Nv_\text{TS}\Sv_\PFt^i\\(\Dv_\TFt-\Gv_\TFt)\Sv_\GFt^i \ebmatrix,
\end{align*}
where $\Sv_\GFt^i$ and $\Sv_\PFt^i$ are the known incident fields applied to the two surfaces. This set of algebraic equations can be solved for the unknown currents $\Cv_\Pt$ and $\Cv_\Gt$; and subsequently, the scattered surface fields $\Sv^\st_\GFt$ and $\Sv^\st_\PFt$. The system matrix is square as the operators $\Nv_\Ct$, $\Dv_\TFt$ and $\Gv_\TFt$ calculate only the tangential components of the fields and the $\Nv_\CSt$ operator only returns the normal component of the surface currents. This completes the matrix formulation of the IE-GSTC framework.

\section{Results and Discussion}

\begin{figure}[h]
    \centering
       \begin{overpic}[width=\columnwidth,grid=false]{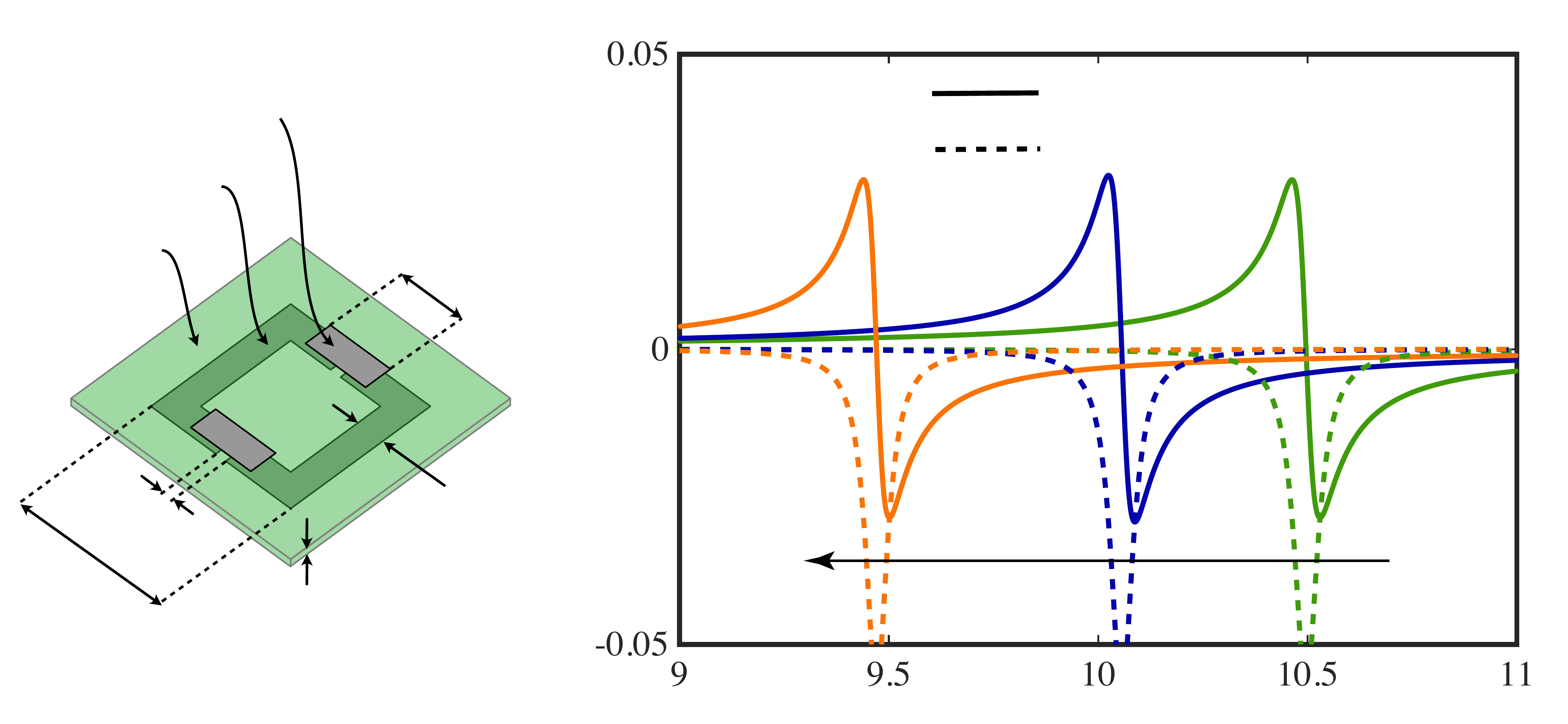}
            \put(68, -1){\scriptsize\htext{Frequency, $f$ (GHz)}}
		    \put(37,23){\scriptsize\vtext{Magnetic Susceptibility, $\chia{mm}{\nn}$}}
		    \put(67, 39.5){\scriptsize $\Re\{\cdot\}$}
		    \put(67, 36){\scriptsize $\Im\{\cdot\}$}
		    \put(48, 9.5){\scriptsize\htext{$\ell$}}
		    \put(20,6){\htext{\tiny 0.025}}
			\put(4,8.5){\htext{\tiny 2.5}}
			\put(8,35){\htext{\tiny bottom copper}}
			\put(13,39){\htext{\tiny top copper}}
			\put(6.5,30.5){\htext{\tiny PI film}}
			\put(8.5,12.5){\htext{\tiny 0.2}}
			\put(30,13.5){\htext{\tiny 0.4}}
			\put(29,28){\htext{\tiny $\ell$}}
			\put(4,4){\tiny \textit{Units are mm}}
        \end{overpic}
        \caption{A practical metasurface unit cell, which is a capacitively-loaded loop on a Polyimide (PI) film where in each cell the side length of a MIM capacitor $\ell$ was can be modulated. The extracted normal magnetic surface susceptibilities as a function of capacitor length $\ell$ is shown. The unit cell dimensions are the same as that of Fig.~\ref{fig:PracticalCell}.}\label{fig:ParamCell}
\end{figure}
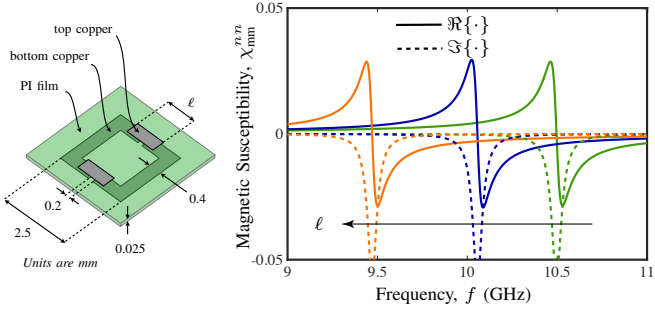

In this section we will consider a practical unit cell based metasurface following the structure of Fig.~\ref{fig:PracticalCell}, which operates around X-band microwave frequencies, and exhibits a strong normal magnetic polarization. Full-wave simulations using Ansys FEM-HFSS were performed using the unit cell with periodic boundary conditions and using plane-wave incidence at two different angles as prescribed in \cite{Xiao:2019aa,VilleFloq}, which revealed a small tangential $\chi_\text{ee}^\zz$ in addition to a strong normal $\chi_\text{mm}^\nn$, as also illustrated in Fig.~\ref{fig:PracticalCell}(b). Furthermore, the surface characteristics can be conveniently controlled by parametrizing the side length $\ell$ of the MIM capacitors, as shown in Fig.~\ref{fig:ParamCell}(a), which very effectively tunes the normal magnetic polarization, $\chi_\text{mm}^\nn$, in particular. The surface susceptibilities were next extracted by curve-fitting for different lengths of the MIM capacitor $\ell$ to build a look-up table, as shown in Fig.~\ref{fig:ParamCell}(b), which can be used to create both uniform and modulated metasurfaces. This surface will now be used to demonstrate the accuracy of the BEM based IE-GSTC simulations using several illustrative examples. 

\subsection{Infinite Uniform Surface}

The first test of the BEM methodology is a uniform surface formed using a constant MIM capacitor length $\ell$, which corresponds to constant surface susceptibilities ($\chi_\ee^\zz$ and $\chi_\mm^\nn$) across the surface at the chosen operation frequency (10~GHz, here). The surface is excited with a uniform plane wave at a fixed angle $\theta$. These values can be used in \eqref{eq:AnaTR} to analytically determine the reflection and transmission values of the surface to validate the BEM computed response. To emulate an infinite surface in BEM, 20 physical cells long metasurface is periodically tiled 128 times using the periodic Greens' function of \eqref{Eq:PerGreen}.

\begin{figure}[htbp]
    \centering
       \begin{overpic}[width= \linewidth,grid=false,clip]{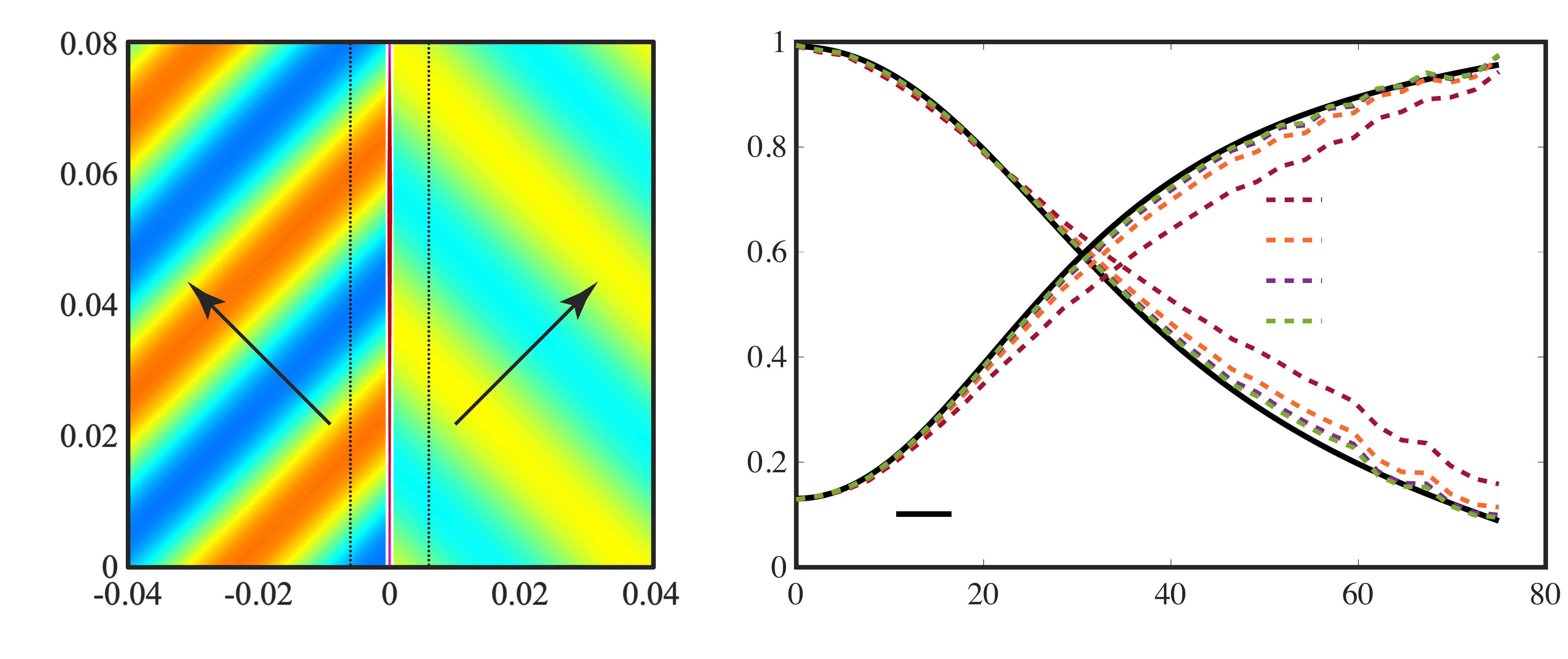}
				\put(25,0){\htext{\scriptsize $x$~(mm)}}
				\put(25, 42){\htext{\scriptsize Scattered $\Re\{E_z^s\}$}}
				\put(75,0){\htext{\scriptsize Incidence Angle $\theta$~(deg)}}
				\put(0,22){\vtext{\scriptsize $y$~(mm)}}
				\put(45,22){\vtext{\scriptsize Scattered Field, $|E_z^{\{\text{r},\text{t}\}}|$}}
				\put(68, 9){\htext{{\tiny Analytical, Eq.~\ref{eq:AnaTR}}}}
				\put(90, 29){\htext{{\tiny $\Delta_\lambda = 5$}}}
				\put(90, 26.5){\htext{{\tiny $\Delta_\lambda = 10$}}}
				\put(90, 24){\htext{{\tiny $\Delta_\lambda = 20$}}}
				\put(90, 21.5){\htext{{\tiny $\Delta_\lambda = 40$}}}
			\end{overpic}
        	\caption{Computed scattered fields from an infinite uniform metasurface at 10~GHz, when incident plane wave launched at $\theta_{i}=\SI{35}{\degree}$, and the comparison of the analytical \eqref{eq:AnaTR} and computed BEM results for various levels of surface discretization as a function of the angle of incidence, $\theta$ and various surface meshing. The BEM simulations use a discretization of $\lambda/40$ in the 2D field plot case and $\chi_\ee^\zz = 0.0013$ and $\chi_\mm^\nn = 0.0241 - j0.0131$.}	\label{fig:UniformSurface}
\end{figure}
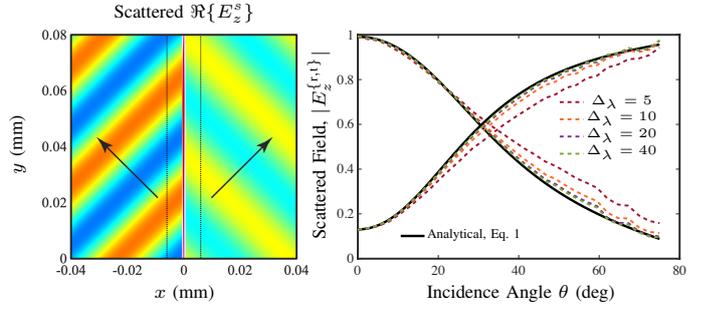

To test the BEM methodology over a large range of incident angles, simulations were performed from normal ($\theta = 0^\circ$) incidence to $\theta = 75^\circ$ and the corresponding scattered fields are computed in reflection and transmission region. The 2D computed E-fields are shown, for instance, in Fig.~\ref{fig:UniformSurface}(a), for an incident angle of 45$^\circ$. As the BEM method employs a discretized surface, the BEM using a full susceptibility description was used with varying levels of discretization. The surface divisions per wavelength ($\Delta_\lambda$) was varied from 5 to 30. Typically a discretization of $\Delta_\lambda = 10-15$ is sufficient for surfaces such a dielectric interfaces or a PEC. However, for the metasurface with a strong normal magnetic polarization component, proper numerical evaluation of surface gradient of the normal component of $\M$ has an impact. For higher angles of incidence, the rapid phase change across the surface requires a larger degree of discretization to correctly capture the response. However, with still a moderate $\Delta_\lambda = 30$, the BEM successfully matches the analytical results of \eqref{eq:AnaTR} well throughout the full range of angles. This further establishes the usefulness of the periodic Greens' function to model infinite surfaces using numerical finite surface.

\subsection{Finite Sized Uniform Metasurface}

Next, consider a finite-sized uniform metasurface, which is excited by a cylindrical wave at \SI{10}{GHz} originating from $(x_0,0)$ with $x_0=-\SI{15}{mm}$ and incident on a variety of uniform surfaces using unit cells with different capacitor lengths. The surface has a length of \SI{0.8}{m}, and on either side of this surface is an implicit dielectric extension.

\begin{figure*}[htbp]
    \centering
       \begin{overpic}[width=2\columnwidth,grid=false,clip]{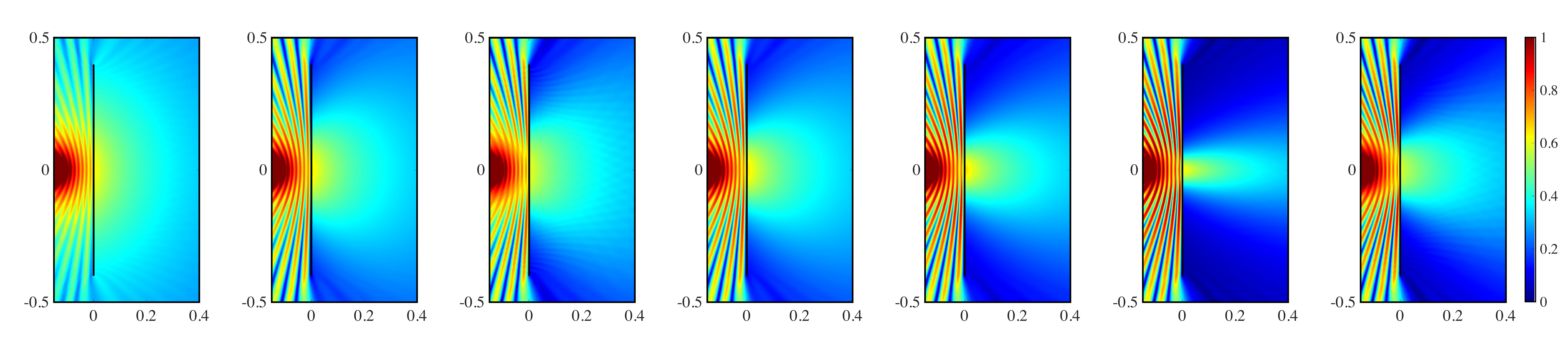}
				\put(8,0){\htext{\scriptsize $x$~(m)}}\put(22,0){\htext{\scriptsize $x$~(m)}}\put(36,0){\htext{\scriptsize $x$~(m)}}\put(50,0){\htext{\scriptsize $x$~(m)}}\put(64,0){\htext{\scriptsize $x$~(m)}}\put(77,0){\htext{\scriptsize $x$~(m)}}\put(91,0){\htext{\scriptsize $x$~(m)}}				
				\put(0,11){\vtext{\scriptsize $y$~(m)}}	\put(14,11){\vtext{\scriptsize $y$~(m)}}\put(28,11){\vtext{\scriptsize $y$~(m)}}\put(42,11){\vtext{\scriptsize $y$~(m)}}\put(56,11){\vtext{\scriptsize $y$~(m)}}\put(70,11){\vtext{\scriptsize $y$~(m)}}\put(84,11){\vtext{\scriptsize $y$~(m)}}
				\put(8,21){\htext{\scriptsize $\chi_\mm^{\nn}= 0$}}\put(22,21){\htext{\scriptsize $\ell = 1.20$}}\put(36,21){\htext{\scriptsize $\ell = 1.22$}}\put(50,21){\htext{\scriptsize $\ell = 1.24$}}\put(64,21){\htext{\scriptsize $\ell = 1.26$}}\put(78,21){\htext{\scriptsize $\ell = 1.28$}}\put(92,21){\htext{\scriptsize $\ell = 1.30$}}
			\end{overpic}           
	\caption{Scattered fields from a finite sized metasurface when excited by a cylindrical source. The total fields $|E_z^\text{t}|$ for a base line simulation for $\chi_\mm^\nn = 0$ (a) and six surfaces with an increasing capacitor length $l = \{1.20,1.22,1.24,1.26,1.28,1.30 \}$ (mm) are shown in (b-g). }
	\label{fig:CylFields2D}
\end{figure*}
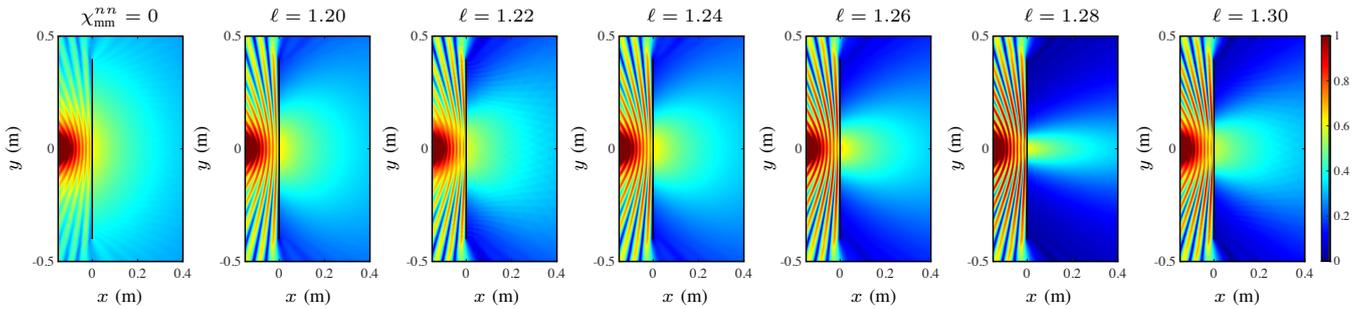

If the incident cylindrical wave is normalized such that $E_z^i(0,0) = 1$ at the center of the surface, the incident fields can be given by:
\begin{align*}
    \E^i_z(x,y) &= \frac{M_f\omega\mu_0}{4}H_{0}^{(2)}\left(k\sqrt{(x-x_0)^2 + y^2 }\right)\\
    \H^i_x(x,y) &= \frac{-jkM_f y}{4\sqrt{(x-x_0)^2} + y^2}H_{1}^{(2)}\left(k\sqrt{(x-x_0)^2 + y^2}\right)\\
    \H^i_y(x,y) &= \frac{-jkM_f\left(x-x_0\right)}{4\sqrt{(x-x_0)^2} + y^2}H_{1}^{(2)}\left(k\sqrt{(x-x_0)^2 + y^2}\right),
\end{align*}
where $H_\nu^{(2)}(X)$ is the Hankel function of the second kind and $M_f = 4/\{\omega\mu_0 H_0^{(2)}(kz_0)\}$ is a normalization factor. The scattered fields from the surface when there is no normal component (i.e. $\chi_\mm^\nn=0$) is shown for reference in Fig. \ref{fig:CylFields2D}(a). Only a small reflection of the incident wave ($R = 0.16$) is observed and the bulk of the wave is passed through the surface. Strong edge diffraction is clearly visible near both edges of the surface.

Next, the normal surface polarization component via $\chi_\mm^\nn$ is introduced. In the central region of the metasurface ($|y|=0$), the fields locally are  at near normal incidence on the surface, so that the surface interaction is dominantly through the tangential susceptibilities, thereby exhibiting large local field transmission. However, for fields near the edges of the surface, the $H_x$ component of the incident field becomes stronger, significantly exciting the magnetic polarization and inducing a stronger reflection. At the edge of the surface the local angle of incidence is $\approx 75^\circ$ with $R = 0.92$ and transmission is almost completely suppressed. A transmission field profile with the strongest intensity near the surface centre which gradually diminishes towards either edge, is clearly seen in  Fig.~\ref{fig:CylFields2D}(b-g), for different strengths of the $\chi_\mm^\nn$ corresponding to larger MIM capacitor length $\ell$. Moreover, the overall transmission across the surface constantly decreases as the normal component $\chi_\mm^\nn$ is increased up to $\ell = 1.26$~mm length, which also manifests as a spatially narrowed transmission beam. Further increasing of the capacitor length decreases the magnitude of the magnetic moment and the transmitted field width starts to broaden again. This example thus further emphasize the importance of the normal surface polarizations in determining the angular response of the surface.

\begin{figure}[htbp]
    \centering
    \begin{subfigure}[htbp]{\columnwidth}
       \begin{overpic}[width= \linewidth,grid=false,clip]{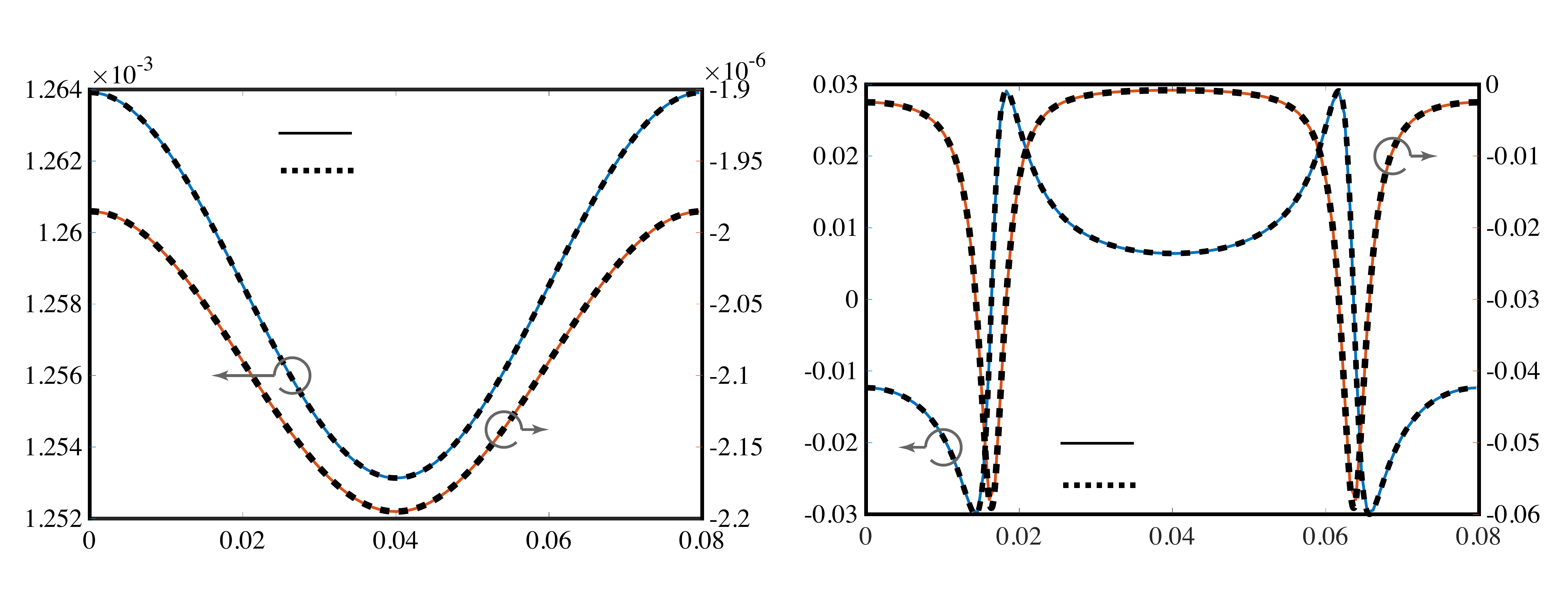}
			\put(25, 0){\htext{\scriptsize distance, $y$ (m)}}
				\put(75, 0){\htext{\scriptsize distance, $y$ (m)}}
				\put(29, 29.5){\htext{ \tiny Floquet, \cite{VilleFloq}}}
				\put(29, 27){\htext{ \tiny BEM-GSTC}}
				\put(78, 10){\htext{ \tiny Floquet, \cite{VilleFloq}}}
				\put(78, 7){\htext{ \tiny BEM-GSTC}}
				\put(75, 36){\htext{\scriptsize \textbf{Normal} $\chi_\ee^{\nn}$ (m)}}
				\put(25, 36){\htext{\scriptsize \textbf{Tangential} $\chi_\ee^{\zz}$ (m)}}
				\put(39, 10){\htext{\scriptsize $\Im\{\cdot\}$}}
				\put(10, 14){\htext{\scriptsize $\Re\{\cdot\}$}}
			\end{overpic}\caption{}
	 \end{subfigure}\hfill
 	\begin{subfigure}[htbp]{\columnwidth}
  	\begin{overpic}[width=\linewidth,grid=false,clip]{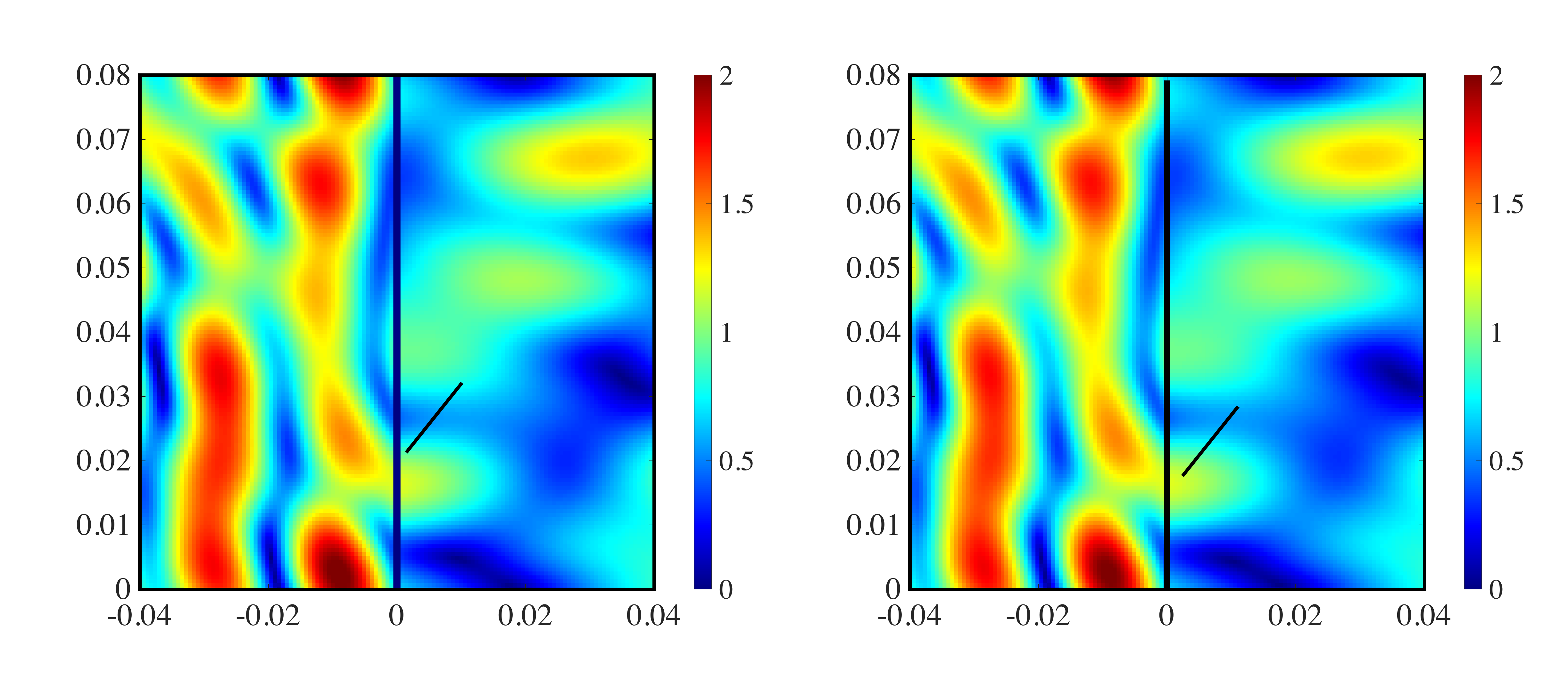}
			\put(75, 1){\htext{\scriptsize distance, $y$ (m)}}
			\put(25, 1){\htext{\scriptsize distance, $y$ (m)}}
			\put(2, 23){\vtext{\scriptsize distance, $x$ (m)}}
			\put(51.5, 23){\vtext{\scriptsize distance, $x$ (m)}}
			\put(75, 42){\htext{\scriptsize \textbf{Floquet Method \cite{VilleFloq}} }}
			\put(25, 42){\htext{\scriptsize \textbf{BEM-GSTC}}}
			\put(83, 21){\htext{\scriptsize Metasurface}}
			\put(33, 21){\htext{\scriptsize Metasurface}}
			\end{overpic}\caption{}
    \end{subfigure}  \hfill
	\begin{subfigure}[htbp]{\columnwidth}
  	\begin{overpic}[width= \linewidth,grid=false,clip]{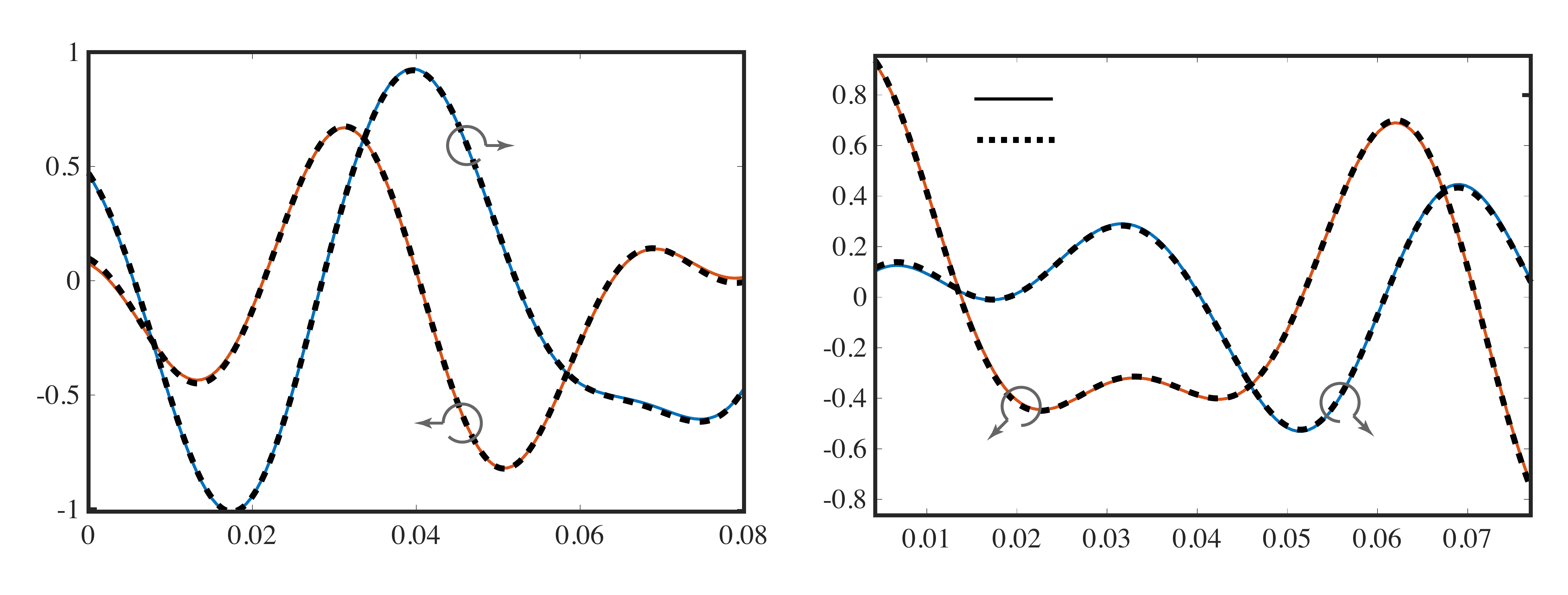}
				\put(25, 0){\htext{\scriptsize distance, $y$ (m)}}
				\put(75, 0){\htext{\scriptsize distance, $y$ (m)}}
				\put(0,20){\vtext{\scriptsize Transmission, $E^{\text{tran.}}_z$}}
				\put(50,20){\vtext{\scriptsize Reflection, $E^{\text{ref.}}_z$}}
				\put(37, 28){\htext{\scriptsize $\Re\{\cdot\}$}}
				\put(23, 10){\htext{\scriptsize $\Im\{\cdot\}$}}
    				\put(73, 31){\htext{ \tiny Floquet, \cite{VilleFloq}}}
				\put(73, 28.5){\htext{ \tiny BEM-GSTC}}
				\put(91, 8){\htext{\scriptsize $\Re\{\cdot\}$}}
				\put(61, 7){\htext{\scriptsize $\Im\{\cdot\}$}}
			\end{overpic}\caption{}
    \end{subfigure}\hfill   
    \begin{subfigure}[htbp]{\columnwidth}
       \begin{overpic}[width= \linewidth,grid=false,clip]{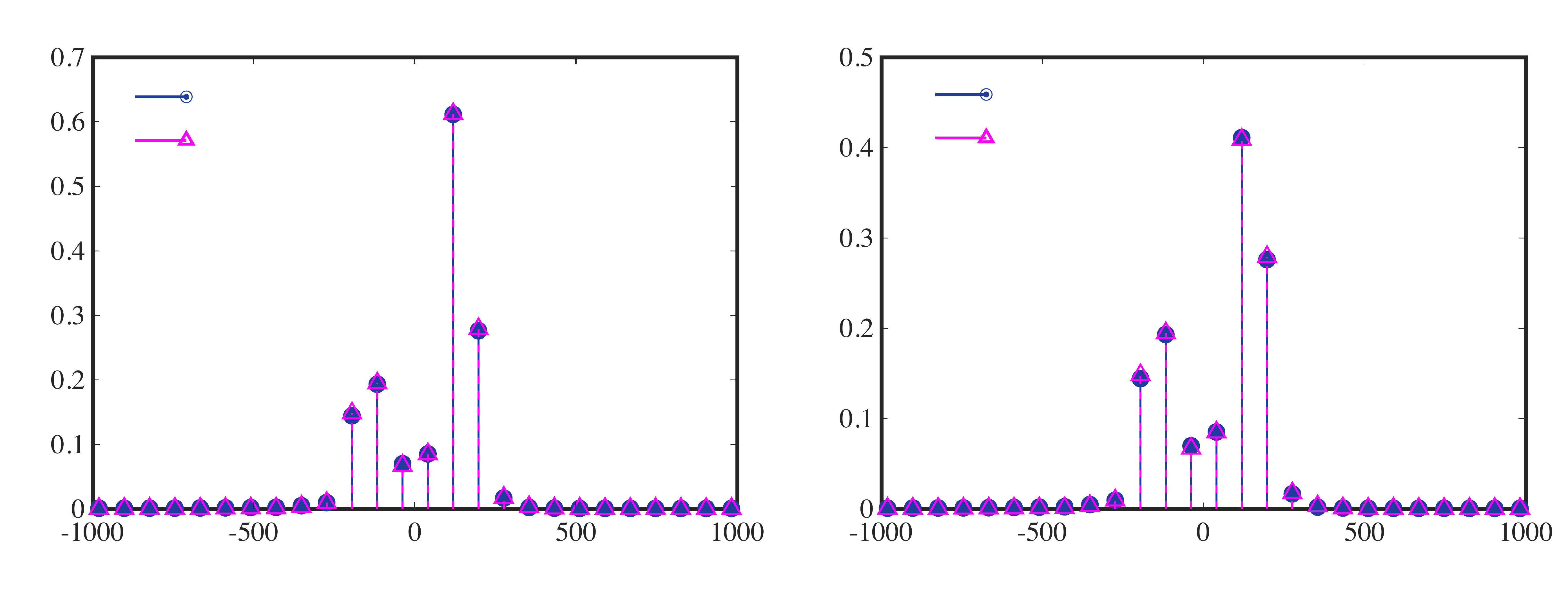}
	  		\put(29, 32.5){\htext{\tiny A}}
	  		    \put(24,15){\htext{\tiny B}}
			    \put(79, 31.5){\htext{\tiny A}}
	  		    \put(74.5,18.5){\htext{\tiny B}}
				\put(26,1){\htext{\scriptsize Spatial Frequency, $k_y$}}
				\put(76,1){\htext{\scriptsize Spatial Frequency, $k_y$}}
				\put(0,20){\vtext{\scriptsize Reflection, $|E_r|/E_{0}$}}
				\put(51,20){\vtext{\scriptsize Tranmission, $|E_t|/E_{0}$}}
				\put(18, 32){\htext{\tiny Floquet, \cite{VilleFloq}}}
				\put(18, 29){\htext{\tiny BEM-GSTC}}
				\put(70, 32){\htext{\tiny Floquet, \cite{VilleFloq}}}
				\put(70, 29){\htext{\tiny BEM-GSTC}}
			\end{overpic}\caption{}
	 \end{subfigure}
	\caption{A periodically modulated metasurface. a) Tangential electric susceptibility: $\chi_\ee^\zz$ and b) Normal magnetic susceptibility: $\chi_\mm^\nn$ profile. c) Scattered E-fields using c) Floquet method with $M=400$ harmonics and d) IE-GSTC using $\lambda/40$ discretization. Fields captures along $x= \pm\lambda/5$ shown in e) and f), for Floquet and IE-GSTCs, respectively. Their spatial Fourier transforms are shown in g) and h). Two harmonics corresponding to the fundamental and a third negative harmonic are labeled `A' and `B' for future reference. Modulation strength used is $\Delta_l = 0.05$.}
	\label{fig:SpatialLoop}
\end{figure}
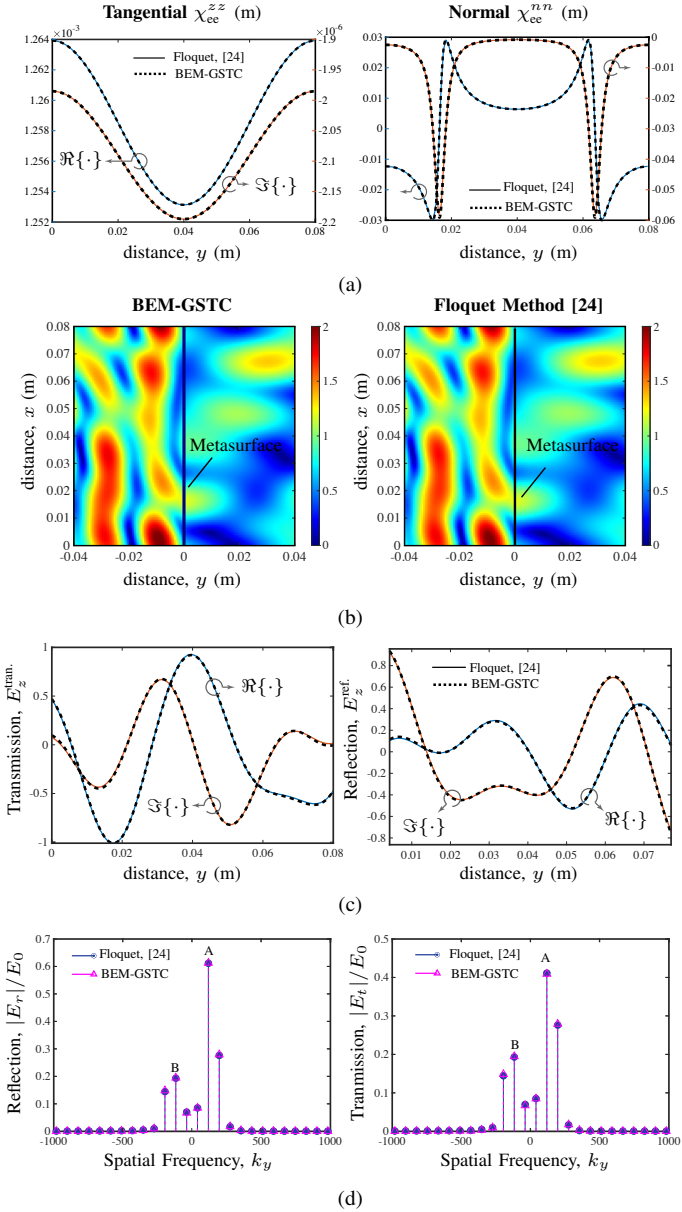
\subsection{Periodically Modulated (Non-uniform) Metasurface}

To increase the complexity of the metasurface, a periodic spatial modulation is next introduced on the surface. A periodic modulation of the MIM capacitor length is chosen as \mbox{$l(y)= \SI{1.25}{mm}+ \Delta_l\cos(\beta_p y)$}, with a length modulation $\Delta_l = \SI{0.05}{mm}$ as the default case and $\beta_p$ as the spatial modulation. Such a periodic surface can be conveniently modeled by a semi-analytical Floquet based field expansion method, which has been shown to capture the metasurface response very accurately with confirmation against full-wave FEM-HFSS simulations \cite{VilleFloq}. A brief summary of the method for the relevant spatial modulation is provided in the Appendix for self-consistency of this work. This Floquet method will be used for independent validation of the BEM results.  

For a periodically varying MIM length $\ell$, the extracted susceptibilities were parameterized and used to create a model of the metasurface. Fig.~\ref{fig:SpatialLoop}(a) and (b) presents both $\chi^\zz_\ee$ and $\chi^\nn_\mm$ along a single period of length $2\pi/\beta_p$. For the BEM simulation the surface was discretized into uniform segments each of a length of $\lambda/40$ and the periodic Greens' function of \eqref{Eq:PerGreen} was used. It should be noted that this level of discretization is primarily determined by the need to accurately represent the variation in the susceptibilities along the surface.

Figures~\ref{fig:SpatialLoop}(b-c) present a comparison between the scattered fields generated by the metasurface obtained using analytical Floquet solution and the computed BEM simulation for an angle of incidence of 35$^\circ$. The modulation of the surface produces a complex field pattern with a large amount of spatial harmonics present. As the surface is periodic and modulated, large number of harmonics are generated in the reflected and transmitted field regions. These, of course, are naturally contained in the Floquet solution, and can be extracted from the BEM simulations by obtaining the fields on vertical lines ($x = \pm x_h$) and then using a spatial Fourier Transform. The accuracy of the BEM method can therefore be evaluated by comparing the fields either in the space domain or in the spatial frequency domain, $k_y$.  A very good agreement is seen in Fig.~\ref{fig:SpatialLoop}(b) between the two based on visual inspection. For closer comparison, the fields are measured along $x = \pm \lambda/5$, in both cases and compared. The transmitted and reflected scattered fields are shown in Figures~\ref{fig:SpatialLoop}(c). The match is excellent and the curves are essentially indistinguishable. These fields can alternatively be decomposed into a set of plane-wave harmonics using a spatial Fourier Transform. This comparison is shown in Fig.~\ref{fig:SpatialLoop}(d) and the match is naturally excellent as well.

\begin{figure}[htbp]
    \centering
     	\begin{subfigure}[htbp]{\columnwidth}
  	\begin{overpic}[width= \columnwidth,grid=false,clip]{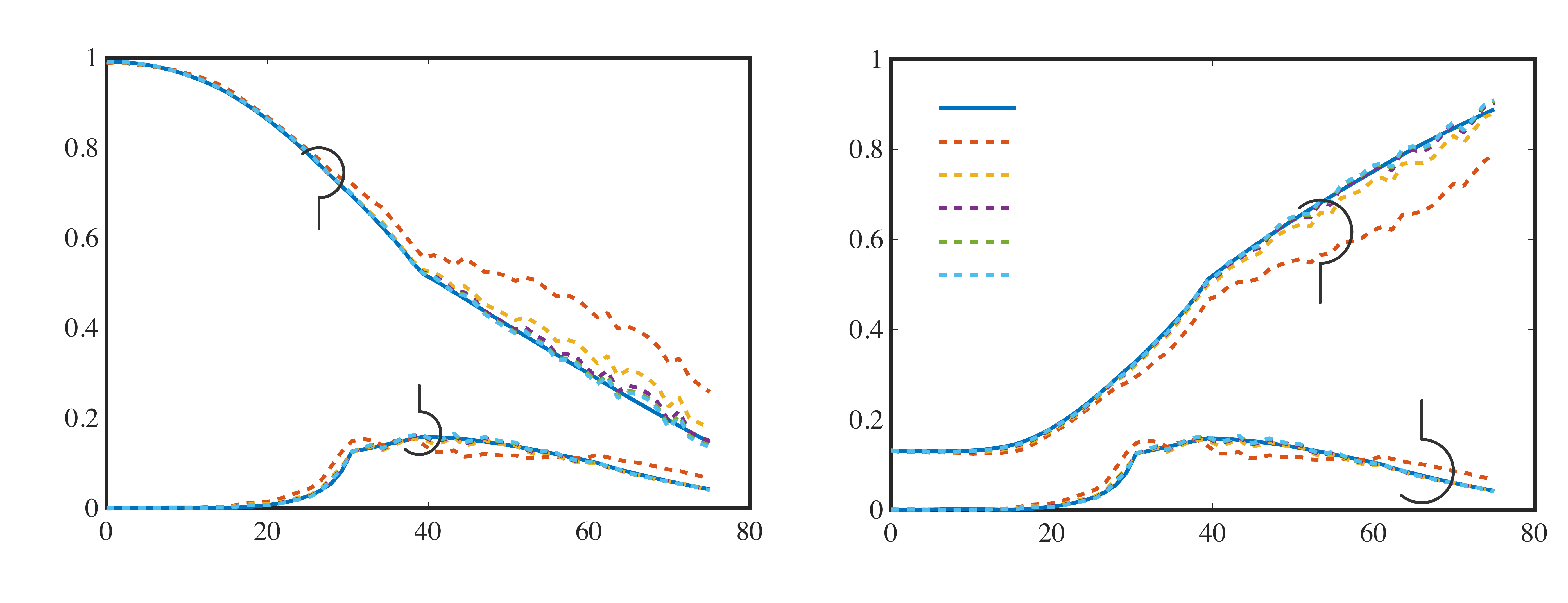}
	  			\put(28, 0){\htext{\scriptsize Incidence Angle, $\theta$ (deg)}}
				\put(78, 0){\htext{\scriptsize Incidence Angle, $\theta$ (deg)}}
				\put(1, 19){\vtext{\scriptsize Reflection Field, $|E^\text{ref.}_z|$}}
				\put(51, 19){\vtext{\scriptsize Transmission Field, $|E^\text{tran.}_z|$}}
				\put(19, 21){\htext{\tiny Harmonic (A)}}
				\put(23, 15){\htext{\tiny Harmonic (B)}}
				\put(87, 17){\htext{\tiny Harmonic (A)}}
				\put(89, 14){\htext{\tiny Harmonic (B)}}
				\put(71, 31){\htext{\tiny Floquet, \cite{VilleFloq}}}
				\put(71, 29){\htext{\tiny $\Delta_\lambda = 5$}}
				\put(71, 27){\htext{\tiny $\Delta_\lambda = 10$}}
				\put(71, 25){\htext{\tiny $\Delta_\lambda = 20$}}
				\put(71, 23){\htext{\tiny $\Delta_\lambda = 40$}}
				\put(71, 21){\htext{\tiny $\Delta_\lambda = 80$}}
			\end{overpic}\caption{}
    \end{subfigure}
     	\begin{subfigure}[htbp]{\columnwidth}
  	\begin{overpic}[width= \columnwidth,grid=false,clip]{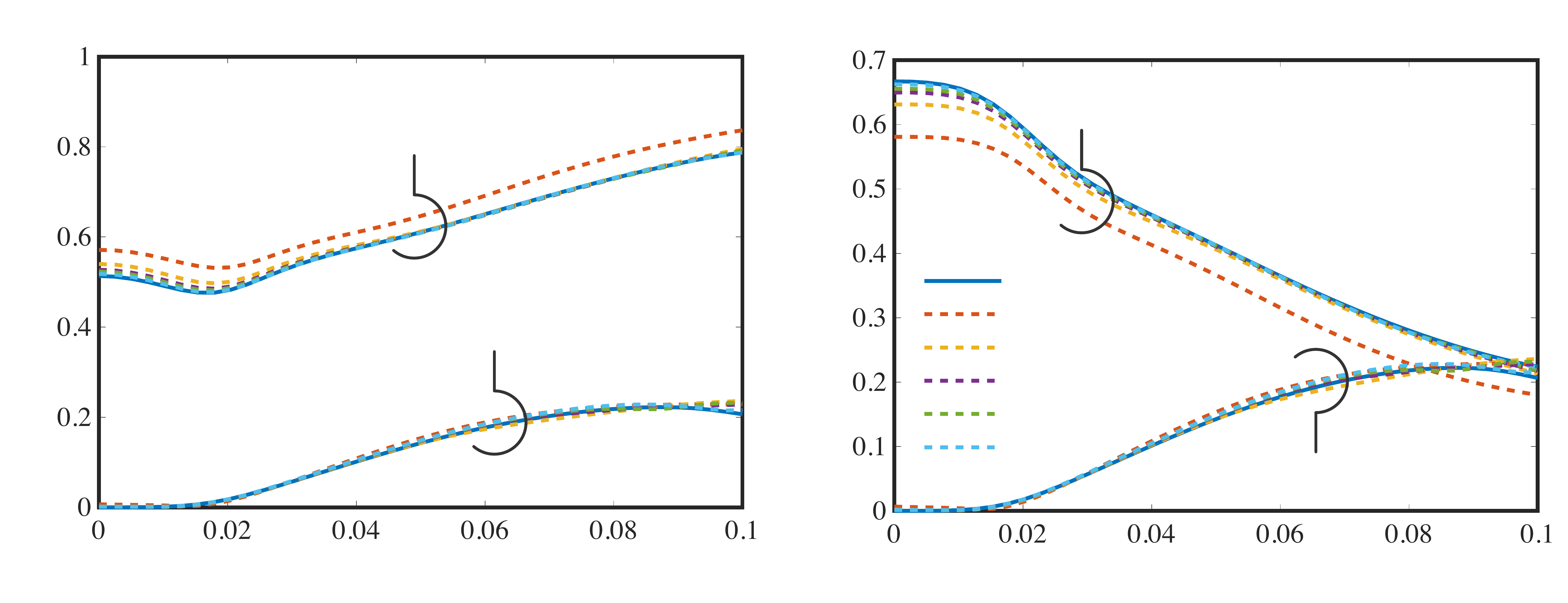}
	  			\put(28, 0){\htext{\scriptsize Modulation Strength, $\l_\Delta$}}
				\put(78, 0){\htext{\scriptsize Modulation Strength, $\l_\Delta$}}
				\put(1, 19){\vtext{\scriptsize Reflection Field, $|E^\text{ref.}_z|$}}
				\put(51, 19){\vtext{\scriptsize Transmission Field, $|E^\text{tran.}_z|$}}
				\put(28, 30){\htext{\tiny Harmonic (A)}}
				\put(35, 17){\htext{\tiny Harmonic (B)}}
				\put(73, 32){\htext{\tiny Harmonic (A)}}
				\put(82, 7.5){\htext{\tiny Harmonic (B)}}
				\put(70, 20){\htext{\tiny Floquet, \cite{VilleFloq}}}
				\put(70, 18){\htext{\tiny $\Delta_\lambda = 5$}}
				\put(70, 16){\htext{\tiny $\Delta_\lambda = 10$}}
				\put(70, 14){\htext{\tiny $\Delta_\lambda = 20$}}
				\put(70, 12){\htext{\tiny $\Delta_\lambda = 40$}}
				\put(70, 10){\htext{\tiny $\Delta_\lambda = 80$}}
			\end{overpic}\caption{}
    \end{subfigure}
	\caption{Two scattered field harmonics corresponding to the fundamental (A) and a third negative harmonic (B), shown in Fig.~\ref{fig:SpatialLoop}, are (a) plotted as function of the angle of incidence of the excitation for a fixed modulation strength of $\Delta_l = 0.05$, and (b) for varying modulation strengths $\Delta_l$ but at a fixed plane wave incidence angle of $35^\circ$. Surface discretization in BEM is varied as $\Delta_\lambda= \{5,10,20,40,80\}$. }
	\label{fig:ModHarm}
\end{figure}

More simulations were undertaken for a variety of incidence angles and modulation strengths. As with the uniform surface the effect of surface discretization $\Delta_\lambda$ was investigated for the modulated surface. This was done with respect to two variables: 1) the angle of incidence of the excitation ($\theta$), and 2) the strength of the modulation ($l_\Delta$). The comparison is not as simple as for the uniform case as we wish to compare not only the overall reflection and transmission characteristics, but the details of the harmonic content. To enable this comparison the magnitude of two harmonics --  the fundamental and an arbitrarily chosen propagating third harmonic (labeled A and B in Fig. \ref{fig:SpatialLoop}) are plotted. Figure \ref{fig:ModHarm}(a-b) shows the comparison of the two methods with respect to the variation of the angle of incidence. The Floquet results are plotted ($M = 400$) and 4 sets of BEM data are presented for increasing discretization. It is clearly seen that the BEM results are converging towards the Floquet results with a smaller discretization needed at higher angles -- this is due to the more rapid phase variation across the surface at the higher angles. This trend is manifested in both the reflected and transmitted fields, where both the fundamental and third harmonic being well predicted. Finally, simulations were also run for a fixed angle of incidence 35$^\circ$ and a range of modulation strengths from \mbox{$l_\Delta = 0.0$--$0.1$}. These results are presented in Fig.~\ref{fig:ModHarm}(c-d). Again, as with the variation with respect to angle of incidence the BEM matches the Floquet results well. To obtain a good result the discretization was found to be greater than 40 with a larger value needed at higher modulation strengths, to accurately model the rapid transitions in the magnetic susceptibility profile across space.

\begin{figure*}[htbp]
    \centering
\begin{subfigure}[htbp]{1.8\columnwidth}    
  	\begin{overpic}[width= \columnwidth,grid=false,clip]{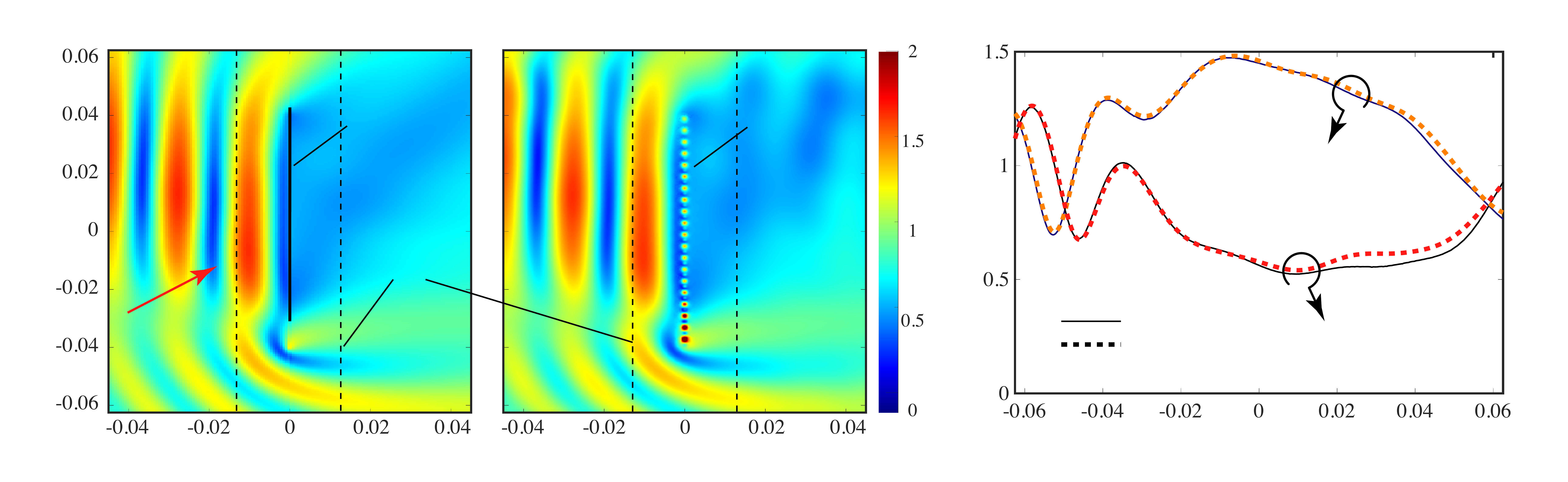}
	  			\put(19 , 1){\htext{\scriptsize Distance, $x$~(m)}}
				\put(43 , 1){\htext{\scriptsize Distance, $x$~(m)}}
				\put(2, 15){\vtext{\scriptsize Distance, $y$~(m)}}
				\put(26 , 24){\htext{\scriptsize \shortstack{$\delta=0$ \\ Metasurface}}}
				\put(50, 24){\htext{\scriptsize \shortstack{$N$-cell \\ Metasurface}}}
				\put(19 , 29){\htext{\scriptsize \color{cobalt}\shortstack{BEM-GSTC}}}
				\put(43, 29){\htext{\scriptsize \color{cobalt}\shortstack{FEM-HFSS}}}
				\put(26 , 14){\htext{\scriptsize \shortstack{Observation \\ Line}}}
				\put(61.5, 15){\vtext{\scriptsize Total Fields, $|E_z^\text{tot.}|$}}
				\put(80 , 2){\htext{\scriptsize Distance, $y$~(m)}}
				\put(75, 10){\htext{\tiny FEM-HFSS}}
				\put(75, 8.5){\htext{\tiny BEM-GSTC}}
				\put(85, 9){\htext{\tiny $x = +0.013$}}
				\put(85, 20){\htext{\tiny $x = -0.013$}}
				\put(-1, 16){\vtext{\small \color{amber}\textsc{Oblique Incidence}, $\theta=+30^\circ$}}
			\end{overpic}\caption{}
\end{subfigure}
\begin{subfigure}[htbp]{1.8\columnwidth}    
  	\begin{overpic}[width= \columnwidth,grid=false,clip]{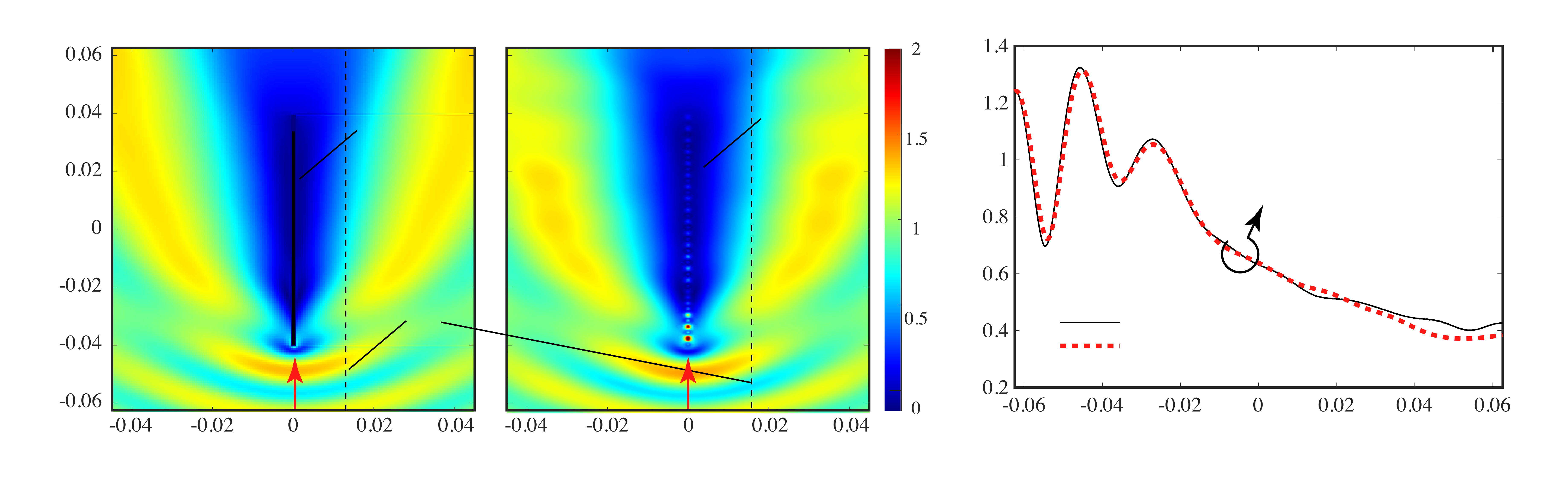}
	  			\put(19 , 1){\htext{\scriptsize Distance, $x$~(m)}}
				\put(43 , 1){\htext{\scriptsize Distance, $x$~(m)}}
				\put(2, 15){\vtext{\scriptsize Distance, $y$~(m)}}
				\put(26 , 24){\htext{\scriptsize \shortstack{$\delta=0$ \\ Metasurface}}}
				\put(52, 24){\htext{\scriptsize \shortstack{$N$-cell \\ Metasurface}}}
				\put(19 , 29){\htext{\scriptsize \color{cobalt}\shortstack{BEM-GSTC}}}
				\put(43, 29){\htext{\scriptsize \color{cobalt}\shortstack{FEM-HFSS}}}
				\put(27 , 12){\htext{\scriptsize \shortstack{Observation \\ Line}}}
				\put(61.5, 15){\vtext{\scriptsize Total Fields, $|E_z^\text{tot.}|$}}
				\put(80 , 2){\htext{\scriptsize Distance, $y$~(m)}}
				\put(75, 10){\htext{\tiny FEM-HFSS}}
				\put(75, 8.5){\htext{\tiny BEM-GSTC}}
				\put(82, 18){\htext{\tiny $x = 0.013$}}
				\put(-1, 16){\vtext{\small \color{amber}\textsc{Grazing Incidence}, $\theta=-90^\circ$}}
			\end{overpic}\caption{}
\end{subfigure}
\begin{subfigure}[htbp]{1.8\columnwidth}    
  	\begin{overpic}[width= \columnwidth,grid=false,clip]{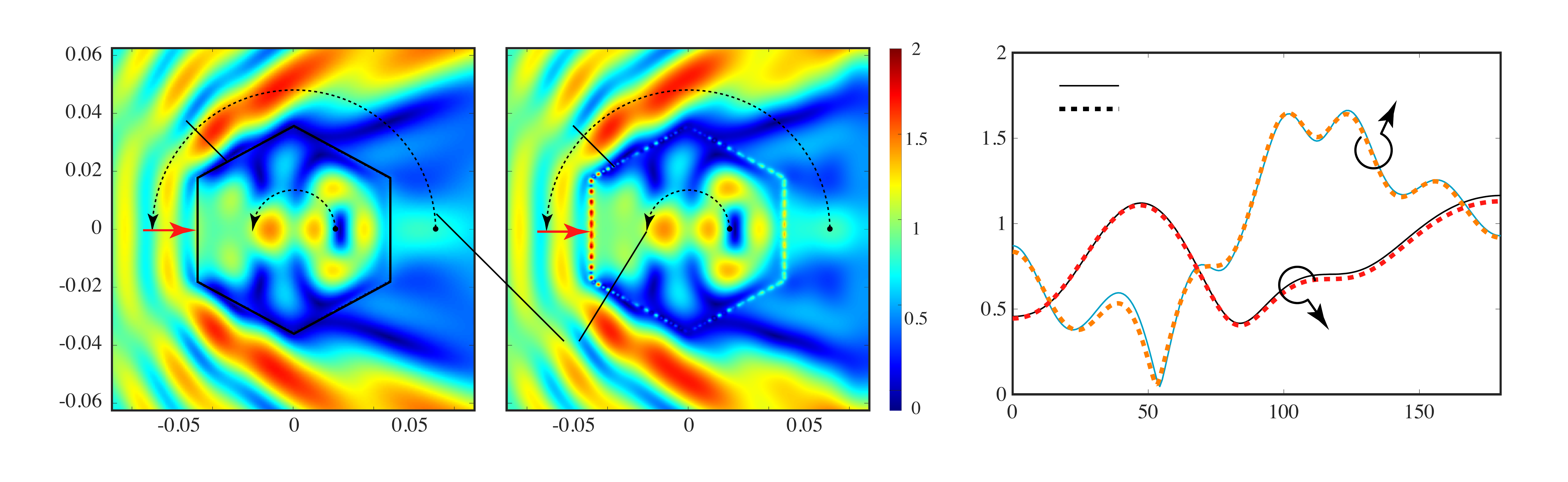}
	  			\put(19 , 1){\htext{\scriptsize Distance, $x$~(m)}}
				\put(43 , 1){\htext{\scriptsize Distance, $x$~(m)}}
				\put(2, 15){\vtext{\scriptsize Distance, $y$~(m)}}
				\put(12 , 24){\htext{\scriptsize \shortstack{$\delta=0$ \\ Metasurface}}}
				\put(37, 24){\htext{\scriptsize \shortstack{$N$-cell \\ Metasurface}}}
				\put(19 , 29){\htext{\scriptsize \color{cobalt}\shortstack{BEM-GSTC}}}
				\put(43, 29){\htext{\scriptsize \color{cobalt}\shortstack{FEM-HFSS}}}
				\put(37 , 6){\htext{\scriptsize \shortstack{Observation \\ Line}}}
				\put(61.5, 15){\vtext{\scriptsize Total Fields, $|E_z^\text{tot.}|$}}
				\put(80 , 2){\htext{\scriptsize Distance, $y$~(m)}}
				\put(75, 25){\htext{\tiny FEM-HFSS}}
				\put(75, 23.5){\htext{\tiny BEM-GSTC}}
				\put(90, 25){\htext{\tiny $r_0 = 0.05$}}
				\put(85, 9){\htext{\tiny $r_0 = 0.015$}}
				\put(-1, 16){\vtext{\small \color{amber}\textsc{Non-Cartesian Surface}}}
			\end{overpic}\caption{}
\end{subfigure}
	\caption{Comparison of the total fields generated by a practical metasurface built using the unit cell of Fig.~\ref{fig:ParamCell} computed using FEM-HFSS and its equivalent zero thickness sheet model computed using the proposed BEM-GSTC framework. a) Finite flat metasurface with oblique incidence. b) Finite flat metasurface with grazing angle incidence. c) Hexagon-shaped non-cartesian metasurface with normal incidence. Simulation parameters are: BEM simulations use a discretization of $\lambda/40$ with $\chi_\ee^\zz = 0.0014$ and $\chi_\mm^\nn =  0.0254 - j0.0159$, corresponding to a capacitor length of $l = \SI{1.25}{mm}$. The two flat surfaces are $\SI{0.08}{m}$ long and the sides of the hexagon are $\SI{0.04}{m}$. The excitation frequency was \SI{10}{GHz}.
	}
	\label{fig:VH_HFSS}
\end{figure*}
\subsection{Zero-thickness Model of a Finite-Thickness Metasurfaces}

So far all the surfaces have been placed along one of the cartesian co-ordinates ($y-$axis) only, while the proposed framework is capable of handling surfaces along arbitrary contours utilizing the local co-ordinate formulation. Moreover, one may also wonder if the zero thickness sheet model described by the tensorial surface susceptibilities is a good representation of a finite thickness and large practical metasurface (and not just the unit cells). To investigate these two aspect and test the BEM-GSTC formulation in the local-co-ordinate system, let us consider a finite sized metasurface built using the unit cell of Fig.~\ref{fig:ParamCell} using some $N$ number of unit cells in FEM-HFSS.

First consider a flat finite sized surface along the $y-$axis, which is excited with a uniform plane-wave at an oblique angle of incidence, as shown in Fig.~\ref{fig:VH_HFSS}(a). It shows the total fields generated by the metasurface in the reflection and transmission regions, using BEM-GSTC and Ansys FEM-HFSS, respectively. An excellent agreement is seen between the two, where BEM-GSTC is able to reproduce fine features of the fields around the surface, notably around the edges. The granularity of the fields near the metasurface is clearly visible in FEM-HFSS, which naturally is not present in the zero-thickness model, as surface susceptibilities describe only the macroscopic field behaviour of the surface. Closer examination of the fields in the transmission and reflection regions slightly away from the surface confirms this agreement as further shown in Fig.~\ref{fig:VH_HFSS}(a), along an observation line. Another example is shown in Fig.~\ref{fig:VH_HFSS}(b), where the metasurface is excited at a grazing angle of incidence. Again an excellent agreement is seen between BEM-GSTCs and FEM-HFSS in the entire region around the surface except the microscopic fields as expected. Due to the symmetry of the fields, comparison between the fields in the transmission region only, is shown along the observation line slightly away from the metasurface.

Let us test a more complicated configuration, such as the one shown in Fig.~\ref{fig:VH_HFSS}(c). The metasurface is shaped into a hexagon, formed using flat uniform metasurface segments forming a lossy cavity. This time the various edges of the structure are not necessarily aligned with the cartesian system, and thus it represents a simple, yet effective, configuration to test the local co-ordinate formulation of BEM-GSTC. The hexagonal structure is excited with a uniform plane wave propagating from the left, normal to the leftmost edge. Excellent agreement is seen between fields generated by the two methods, everywhere inside and outside closed region. Closer examination of the fields along predefined observation paths inside and outside confirm this agreement.
 
 All these results thus confirm that the zero thickness model using appropriately chosen surface susceptibilities is an accurate representation of a finite thickness metasurface producing correct macroscopic fields. While this agreement may intuitively be expected of the surface susceptibilities, the close agreement between the generated fields by BEM-GSTC and FEM-HFSS for various cases above, can still be considered remarkable. Its a clear demonstration of the usefulness of the zero-thickness model to compute field scattering from metasurfaces as opposed to a brute-force 3D full-wave simulation, which naturally comes with a computational advantage requiring less resources. These various examples further successfully validate the proposed BEM-GSTC framework.

\section{Conclusions}

An IE-GSTC field solver using metasurface susceptibility tensors with normal surface polarizabilities has been proposed and validated using variety of numerical examples in 2D. The method solves for scattered fields from the metasurface which are represented as spatial discontinuities and described using surface susceptibilities. It self-consistently computes these fields by solving the GSTCs and IE based field propagation using BEM. Incorporating the complete tensorial surface susceptibilities including the normal polarization components and their spatial derivatives along the surface, the method has been demonstrated to accurately model the angular scattering properties of practical metasurfaces. The field equation formulation utilizing a local co-ordinate system further enables modeling metasurfaces with arbitrary orientations and configurations. Using a variety of examples, the proposed 2D BEM-GSTC framework has been tested and subsequently used to demonstrate that the zero-thickness sheet model with complete tensorial susceptibilities can very accurately reproduce the macroscopic fields of both uniform and periodic surfaces, while accounting for edge diffraction effects of finite sized metasurfaces.

\section*{Acknowledgements}

The authors acknowledge funding from the Department of National Defence's Innovation for Defence Excellence and Security (IDEaS) Program in support of this work.

\section{Appendix} \label{sec:fl}

If the unit cell shown in Fig. \ref{fig:PracticalCell} is used to build a periodic modulated surface with spatial angular frequency $\beta_\text{p}=2\pi/L$ then a Floquet methodology can be used to analyze the field response in the spatial frequency domain \cite{VilleFloq}. This method expresses the fields as,
\begin{align*}
	\E^\text{a}(x,y) 
		&= \sum_{m=-\infty}^{\infty}  E_\text{0,a}(m) e^{j[- k_{y}(m)y \pm j k_{x}(m) x]} \zh\\
		&= \sum_{m=-\infty}^{\infty} E_\text{a}(m,x,y) \zh\\
	\H^{a}(x,y) &= \frac{1}{k\eta_0}\sum_{m=-\infty}^{\infty} E_{a}(m,x,y) 
		\left[k_{y}(m)\xh \pm k_{x}(m)\yh \right]
\end{align*}
where $\text{a}=(\text{i},\text{r},\text{t})$ for the incident, reflected, and transmitted fields, respectively. {The sign on $\pm jk_{z,mn} z$ is ($-$) for incident and transmitted harmonics and ($+$) for reflected harmonics.} Only a single harmonic is present for the incident field ($m_\text{i}$) corresponding to a plane wave. Likewise, we can write the polarization densities, which are
\begin{align*}
	P_z(y) &= \sum_{m=-\infty}^{\infty}  P_{z}(m)e^{-j k_{y}(m)}\\
	M_{\{x,y\}}(x) &= \sum_{m=-\infty}^{\infty} M_{\{x,y\}}(m)e^{-j k_{y}(m)}.
\end{align*}

Floquet's theorem prescribes that the transverse part of the wave-vector ($k_y$) takes on discrete values determined by spatial periodicity, and the normal component ($k_x$) then follows from having a total magnitude $k$:
\begin{align*}
	k_{y}(m) &= k_0\sin\theta_{0} + m\beta_\text{p},\,m\in\text{integers},\\
	k_{x}(m) &= \sqrt{k^2-k_{y}^2(m)},
\end{align*}%
The angles of scattering for the harmonics are found from $\sin[\theta(m)]=k_{y}(m)/k$, which yields
\begin{align*}
	\sin[\theta(m)] = \sin\theta_{0}+m\beta_\text{p}/k_0
\end{align*}
where $\theta_{0}=\theta(0)$ is the angle of the fundamental harmonic. 

There are five sets of unknown harmonic amplitudes, $E_\text{r}$, $E_\text{t}$, $P_z$, $M_x$, and $M_y$, for the TE illumination in consideration. Each set of unknowns has the form,
\begin{align}
	\Ev_\text{t} = \begin{bmatrix}
		\cdots,E_\text{t}(m_{-1}), E_\text{t}(m_{0}), E_\text{t}(m_1), \cdots
	\end{bmatrix}_,^T \notag
\end{align}
and likewise for $\Ev_\text{r}$, $\Pvbb_y$, $\Mv_x$, and $\Mv_y$. We also define matrices with the diagonal terms given by,
\begin{gather*}
    \quad \Kv_x(m,m) = k_{x}(m), \text{and} \quad \Kv_y(m,m)=k_{y}(m).
\end{gather*} 
Substituting the field expansions into the constitutive relations \eqref{eq:ConstitutiveRelations} (while noting that the local coordinate system is the global coordinate system for this flat surface),
\begin{subequations}\label{Eq:MatrixEqnsNoDisp}
	\begin{align}
		\Pvbb_z &= \frac{\epsilon_0}{2}\Cv_\text{e}^{z}(\Ev_\text{r}+\Ev_\text{t}+\Ev_\text{i})\\
		\Mv_y &= \frac{1}{2\eta_0 k}\Cv_\text{m}^{t} \Kv_z (\Ev_\text{r}-\Ev_\text{t}-\Ev_\text{i})\\
		\Mv_x &= \frac{1}{2\eta_0 k}\Cv_\text{m}^{n} \Kv_x (\Ev_\text{r}+\Ev_\text{t}+\Ev_\text{i})
	\end{align}
where $a\oslash b$ is Hadamard (element-wise) division and with
\begin{align*}
	\Cv_{\{\text{e,m}\}}^{\{t,z,n\}}(m_1,m_2) = \chi_{\{\text{ee,mm}\}}^{\{tt,zz,nn\}}(m_1-m_2)
\end{align*}
where
\begin{align}
	\chi_{\{\text{ee,mm}\}}^{\{tt,zz,nn\}}(y) = \sum_{m=-\infty}^\infty \chi_{\{\text{ee,mm}\}}^{\{tt,zz,nn\}}(m) e^{-jm\beta_\text{p} y}.\notag
\end{align}

For the unit cell we are concerned with we have two non-zero susceptibility components $\chi_\ee^\zz$ and $\chi_\mm^\nn$ which are Fourier expanded for use in $\Cv_{\{\text{e,m}\}}^{\{t,z,n\}}$. 

Next, the expansions of the fields and polarization densities are substituted into the GSTC equations \eqref{eq:GSTC}:
%\begin{subequations}
\begin{gather}\textbf{}
		j\mu_0\omega\Mv_y = \Ev_\text{t} - \Ev_\text{r} -\Ev_\text{i}\\
   		j\omega\Pvbb_z - j\Kv_y\Mv_x = \frac{1}{k\eta_0} \Kv_x(-\Ev_\text{r} - \Ev_\text{t}+\Ev_\text{i}) 
\end{gather}\label{eq:Floquet}
\end{subequations}%
Together these equations \eqref{eq:Floquet} provide five sets of equations to solve for the unknown fields and polarizations, given $\Ev_\text{i}$ and parameterized by the susceptibilities. Of course, the number of harmonics must be truncated to $m\in [-M,M]$ to make the problem numerically tractable.

\balance
\bibliographystyle{IEEEtran}
\bibliography{2020_Metasurface_Smy.bib}

% Generated by IEEEtran.bst, version: 1.13 (2008/09/30)
\begin{thebibliography}{10}
\providecommand{\url}[1]{#1}
\csname url@samestyle\endcsname
\providecommand{\newblock}{\relax}
\providecommand{\bibinfo}[2]{#2}
\providecommand{\BIBentrySTDinterwordspacing}{\spaceskip=0pt\relax}
\providecommand{\BIBentryALTinterwordstretchfactor}{4}
\providecommand{\BIBentryALTinterwordspacing}{\spaceskip=\fontdimen2\font plus
\BIBentryALTinterwordstretchfactor\fontdimen3\font minus
  \fontdimen4\font\relax}
\providecommand{\BIBforeignlanguage}[2]{{%
\expandafter\ifx\csname l@#1\endcsname\relax
\typeout{** WARNING: IEEEtran.bst: No hyphenation pattern has been}%
\typeout{** loaded for the language `#1'. Using the pattern for}%
\typeout{** the default language instead.}%
\else
\language=\csname l@#1\endcsname
\fi
#2}}
\providecommand{\BIBdecl}{\relax}
\BIBdecl

\bibitem{MS_review_Yu}
H.-T. Chen, A.~J. Taylor, and N.~Yu, ``A review of metasurfaces: physics and
  applications.'' \emph{Reports on progress in physics. Physical Society},
  vol.~79, no.~7, p. 076401, 2016.

\bibitem{Chi_Review}
X.~Liu, F.~Yang, M.~Li, and S.~Xu, ``Generalized boundary conditions in surface
  electromagnetics: Fundamental theorems and surface characterizations,''
  \emph{Appl. Sci.}, vol.~9, p. 1891, 2019.

\bibitem{GSTC_Holloway}
C.~L. Holloway and E.~F. Kuester, ``Generalized sheet transition conditions for
  a metascreen—a fishnet metasurface,'' \emph{IEEE Trans. Antennas Propag.},
  vol.~66, no.~5, pp. 2414--2427, May 2018.

\bibitem{meta2}
C.~Holloway, E.~F. Kuester, J.~Gordon, J.~O'Hara, J.~Booth, and D.~Smith, ``An
  overview of the theory and applications of metasurfaces: The two-dimensional
  equivalents of metamaterials,'' \emph{IEEE Antennas Propag. Mag.}, vol.~54,
  no.~2, pp. 10--35, April 2012.

\bibitem{MS_Synthesis}
K.~Achouri, M.~A. Salem, and C.~Caloz, ``General metasurface synthesis based on
  susceptibility tensors,'' \emph{IEEE Trans. Antennas Propag.}, vol.~63,
  no.~7, pp. 2977--2991, Jul 2015.

\bibitem{Chi_extraction_Macrodmodel}
U.~R. {Patel}, P.~{Triverio}, and S.~V. {Hum}, ``A fast macromodeling approach
  to efficiently simulate inhomogeneous electromagnetic surfaces,'' \emph{IEEE
  Trans. Antennas Propag.}, pp. 1--1, 2020.

\bibitem{TBC_vs_GSTC_Caloz}
M.~{Dehmollaian}, G.~{Lavigne}, and C.~{Caloz}, ``Comparison of tensor boundary
  conditions with generalized sheet transition conditions,'' \emph{IEEE Trans.
  Antennas Propag.}, vol.~67, no.~12, pp. 7396--7406, Dec 2019.

\bibitem{KuesterGSTC}
E.~F. Kuester, M.~A. Mohamed, M.~Piket-May, and C.~L. Holloway, ``Averaged
  transition conditions for electromagnetic fields at a metafilm,'' \emph{IEEE
  Trans. Antennas Propag.}, vol.~51, no.~10, pp. 2641--2651, Oct 2003.

\bibitem{GenBCEM}
\BIBentryALTinterwordspacing
X.~Liu, F.~Yang, M.~Li, and S.~Xu, ``Generalized boundary conditions in surface
  electromagnetics: Fundamental theorems and surface characterizations,''
  \emph{Applied Sciences}, vol.~9, no.~9, 2019. [Online]. Available:
  \url{https://www.mdpi.com/2076-3417/9/9/1891}
\BIBentrySTDinterwordspacing

\bibitem{Caloz_MS_Siijm}
Y.~Vahabzadeh, N.~Chamanara, K.~Achouri, and C.~Caloz, ``Computational analysis
  of metasurfaces,'' \emph{IEEE Journal on Multiscale and Multiphysics
  Computational Techniques}, vol.~3, pp. 37--49, 2018.

\bibitem{Caloz_Spectral}
N.~Chamanara, K.~Achouri, and C.~Caloz, ``Efficient analysis of metasurfaces in
  terms of spectral-domain gstc integral equations,'' \emph{IEEE Trans.
  Antennas Propag.}, vol.~65, no.~10, pp. 5340--5347, Oct 2017.

\bibitem{Smy_Metasurface_Space_Time}
S.~A. {Stewart}, T.~J. {Smy}, and S.~{Gupta}, ``Finite-difference time-domain
  modeling of space–time-modulated metasurfaces,'' \emph{IEEE Trans. Antennas
  Propag.}, vol.~66, no.~1, pp. 281--292, Jan 2018.

\bibitem{Smy_Close_ILL}
T.~J. {Smy} and S.~{Gupta}, ``Surface susceptibility synthesis of metasurface
  skins/holograms for electromagnetic camouflage/illusions,'' \emph{IEEE
  Access}, vol.~8, pp. 226\,866--226\,886, 2020.

\bibitem{smy2020IllOpen}
T.~J. {Smy}, S.~A. {Stewart}, and S.~{Gupta}, ``Surface susceptibility
  synthesis of metasurface holograms for creating electromagnetic illusions,''
  \emph{IEEE Access}, vol.~8, pp. 93\,408--93\,425, 2020.

\bibitem{stewart2019scattering}
S.~A. {Stewart}, S.~{Moslemi-Tabrizi}, T.~J. {Smy}, and S.~{Gupta},
  ``Scattering field solutions of metasurfaces based on the boundary element
  method for interconnected regions in 2-{D},'' \emph{IEEE Trans. Antennas
  Propag.}, vol.~67, no.~12, pp. 7487--7495, Dec 2019.

\bibitem{FE_BEM_Impedance}
S.~He, W.~E.~I. Sha, L.~Jiang, W.~C.~H. Choy, W.~C. Chew, and Z.~Nie,
  ``Finite-element-based generalized impedance boundary conditions for modeling
  plasmonic nanostructures,'' \emph{IEEE Trans. Nanotechnol.}, pp. 336--345,
  Mar. 2012.

\bibitem{Caloz_MS_IE}
M.~{Dehmollaian}, N.~{Chamanara}, and C.~{Caloz}, ``Wave scattering by a
  cylindrical metasurface cavity of arbitrary cross section: Theory and
  applications,'' \emph{IEEE Trans. Antennas Propag.}, vol.~67, no.~6, pp.
  4059--4072, June 2019.

\bibitem{AppBEMEM}
S.~Kagami and I.~Fukai, ``Application of boundary-element method to
  electromagnetic field problems,'' \emph{IEEE Trans. Microw. Theory Tech.},
  no.~4, pp. 455--461, Apr. 1984.

\bibitem{Smy_EuCap_BEM_2020}
T.~J. Smy, J.~Connor, S.~A. Stewart, and S.~Gupta, ``General formulation of the
  boundary element method (bem) for curvilinear metasurfaces in the presence of
  multiple scattering objects,'' in \emph{14th European Conference on Antennas
  and Propagation (EuCap), Copenhagen, Denmark}, Mar 2020, pp. 1--2.

\bibitem{Caloz_EM_inversion}
T.~{Brown}, C.~{Narendra}, Y.~{Vahabzadeh}, C.~{Caloz}, and P.~{Mojabi}, ``On
  the use of electromagnetic inversion for metasurface design,'' \emph{IEEE
  Trans. Antennas Propag.}, pp. 1--1, 2019.

\bibitem{CalozFDTD}
Y.~Vahabzadeh, K.~Achouri, and C.~Caloz, ``Simulation of metasurfaces in finite
  difference techniques,'' \emph{IEEE Trans. Antennas Propag.}, vol.~64,
  no.~11, pp. 4753--4759, Nov 2016.

\bibitem{Karim_Angular_MS}
K.~{Achouri} and O.~J.~F. {Martin}, ``Angular scattering properties of
  metasurfaces,'' \emph{IEEE Trans. Antennas Propag.}, vol.~68, no.~1, pp.
  432--442, 2020.

\bibitem{Karim_Bianiso_MS}
------, ``Fundamental properties and classification of polarization converting
  bianisotropic metasurfaces,'' \emph{IEEE Trans. Antennas Propag.}, pp. 1--1,
  2021.

\bibitem{VilleFloq}
V.~Tiukuvaara, T.~Smy, and S.~Gupta, ``Floquet analysis of space-time modulated
  metasurfaces with lorentz dispersion,'' \emph{IEEE Trans. Antennas Propag.},
  2021.

\bibitem{Xiao:2019aa}
X.~Liu, F.~Yang, M.~Li, and S.~Xu, ``Generalized boundary conditions in surface
  electromagnetics: Fundamental theorems and surface characterizations,''
  \emph{Appl. Sci.}, vol.~9, no.~9, 2019.

\bibitem{chew2009integral}
W.~Chew, M.~Tong, and B.~Hu, \emph{Integral Equation Methods for
  Electromagnetic and Elastic Waves}.\hskip 1em plus 0.5em minus 0.4em\relax
  Morgan \& Claypool Publishers, 2009.

\bibitem{Method_Moments}
W.~C. Gibson, \emph{The Method of Moments in Electromagnetics}.\hskip 1em plus
  0.5em minus 0.4em\relax Chapman \& Hall, 2008.

\bibitem{linton1998green}
C.~Linton, ``The green's function for the two-dimensional helmholtz equation in
  periodic domains,'' \emph{Journal of Engineering Mathematics}, vol.~33,
  no.~4, pp. 377--401, 1998.

\bibitem{IdemenDiscont}
M.~M. Idemen, \emph{Discontinuities in the Electromagnetic Field}.\hskip 1em
  plus 0.5em minus 0.4em\relax John Wiley \& Sons, 2011.

\end{thebibliography}

\end{document}